\documentclass[aps,prb,reprint,onecolumn,10pt,a4paper]{revtex4-2}
\usepackage[T1]{fontenc}
\usepackage[utf8]{inputenc}
\usepackage{amsmath,amssymb,amsthm,bm,mathtools}
\usepackage{booktabs,array,graphicx,microtype,xcolor}
\newcommand{\rev}[1]{#1}
\usepackage[hidelinks]{hyperref}
\usepackage{orcidlink}
\usepackage{enumitem}
\setlist{nosep}
\newcommand{\ket}[1]{\lvert#1\rangle}
\newcommand{\bra}[1]{\langle#1\rvert}
\newcommand{\Tr}{\operatorname{Tr}}
\newcommand{\diag}{\operatorname{diag}}
\newcommand{\sech}{\operatorname{sech}}
\newcommand{\atanh}{\operatorname{artanh}}
\newcommand{\id}{\mathbb{I}}
\newcommand{\PPT}{\mathrm{PPT}}
\newcommand{\LOCC}{\mathrm{LOCC}}
\newcommand{\NL}{\mathrm{NL}}
\newcommand{\cc}{\mathrm{cc}}
\newcommand{\Mcal}{\mathcal{M}}
\newcommand{\Dcal}{\mathcal{D}}
\newcommand{\hbin}{h_2}
\theoremstyle{definition}
\newtheorem{theorem}{Theorem}[section]
\theoremstyle{plain}
\newtheorem{proposition}[theorem]{Proposition}
\theoremstyle{definition}
\newtheorem{corollary}[theorem]{Corollary}
\theoremstyle{remark}

\numberwithin{equation}{section}
\begin{document}
\color{black}
\title{Entanglement distillation, nonlocal magic, and state sharing\\
in an isospectral non-Hermitian bipartite model}
\author{Chen-Huan Wu \orcidlink{0000-0003-1020-5977}}
\thanks{chenhuanwu1@gmail.com}
\affiliation{Department of Physics, Faculty of Science, Universiti Malaya, Kuala Lumpur 50603, Malaysia}
\begin{abstract}
An isospectral nonunitary similarity deformation changes normalized right
eigenstates without changing a Hamiltonian's eigenvalues. We use a two-qubit
realization to study noisy entanglement distillation, nonlocal magic, and state
sharing while keeping the spectrum fixed. The pure right-eigenstate family is
embedded in a two-parameter maximally correlated family that includes local
dephasing and a calibrated noisy preparation circuit. Its optimized CHSH value
is \(2\sqrt{1+t^2}\), so every member with nonzero coherence violates the CHSH
inequality, whereas complete dephasing produces a classically correlated
separable state. Hashing and a diagonal relative-entropy comparison give the
exact distillable entanglement, every finite-level PPT relative entropy, and
their regularizations. Applying established correlated-catalytic PPT
monotonicity then shows that catalysts cannot improve the Bell-pair
distillation rate. We also derive a model-constrained finite-sample rate
certificate. Further calculations give a Bell participation law,
stabilizer-work profile, Gaussian resource separation, and exact swapping
recursion. Finite-copy spectral certificates specify how the framework extends
beyond one pair. The binomial Schmidt spectrum explains why pairwise optimized
magic is extensive while collective local-unitary optimization leaves
logarithmic growth and a strictly vanishing regularized density, despite
positive entanglement density. Fixed Clifford decoding gives exact six- and
eight-copy CCZ concentration probabilities obeying
\(P_6=\frac12(1-2^{-\mathcal M_\gamma})\) and
\(P_8=1-2^{-\mathcal M_\gamma}\), together with a positive global stabilizer
yield. Explicit replica eigenmodes connect Gaussian-copy invariance to the
fixed-point algebra protecting the original parity support.
Symmetry-resolved activation, equal-fidelity noisy preparations, and a
distance-two fermionic code calculation show how these resources depend on
the allowed operations.
\rev{A complementary stabilizer-fidelity measure is obtained exactly from the
same Schmidt spectrum and grows as one half of the logarithm of the copy
number, while an entropy-subtracted local stabilizer entropy supplies a second
closed resource profile. Random parity twirling and the Bell-endpoint Choi
construction further separate syndrome dephasing, parity postselection, and
entanglement-assisted magic injection.}
\end{abstract}
\maketitle

\section{Introduction}
Eigenvalues do not determine the operational resources of an eigenstate.
In particular, an invertible nonunitary similarity transformation preserves
a Hamiltonian's spectrum while changing its normalized right eigenvectors in
a fixed tensor-product Hilbert space. Here we specify the two-qubit
Hamiltonian and its physical inner product explicitly in
Sec.~\ref{sec:model}. We study how noisy preparation changes distillable
entanglement, how different definitions of magic behave on many copies, and
how the resulting state can be shared or concentrated under specified
operation classes. Related pure-state results are available in the preprint
of Ref.~\cite{Wu}.

The pure right-state family is embedded in the real maximally correlated
family \(\rho(u,t)\), whose coherence \(t\) is attenuated by local phase noise.
Its exact optimized CHSH value is \(2\sqrt{1+t^2}\), while its entanglement
manipulation is controlled by
\begin{equation}
 \Dcal(u,t)=\hbin\!\left(\frac{1+u}{2}\right)
 -\hbin\!\left(\frac{1+\sqrt{u^2+t^2}}{2}\right),
 \label{eq:introD}
\end{equation}
where \(\hbin(z)=-z\log_2z-(1-z)\log_2(1-z)\).
Regularizing LOCC-optimized coherent information yields distillable
entanglement \cite{DevetakWinter,Takagi2026}, whereas regularizing a
relative-entropy distance to a PPT comparison set yields an entanglement
monotone \cite{Ao2026}. These constructions agree here because the states are
maximally correlated: a hashing lower bound and a diagonal comparison state
meet exactly \cite{Rains}. This equality also makes the correlated-catalytic
question tractable without introducing PPT bound entanglement or claiming
general mixed-state reversibility.

The Pauli moments of the pure boundary give the closed nonlocal stabilizer
R\'enyi entropy
\(\Mcal_\gamma=-\log_2(1-u_\gamma^2+u_\gamma^4)\).
Bell participation and stabilizer statistical mechanics provide operational
and generating-function descriptions of these moments
\cite{HoshinoCFT,Salazar}, while a fermionic interpretation exposes
Gaussian/stabilizer resource separation \cite{HoshinoFNG}. Pairwise noise
provides exact protected-support and cross-dimer tests
\cite{FechisinPower}. Heralded state sharing \cite{HuhtanenSwap} supplies a
network application, and rank-limited spectral certificates \cite{Sierant}
keep the assumptions behind collective local-unitary optimization explicit.

A fixed Clifford CNOT decodes the same state into a pure single-qubit
resource, enabling exact universal CCZ concentration
\cite{MagicConcentration}. The model-specific probabilities satisfy
\begin{equation}
 P_6(\gamma)=\tfrac12u_\gamma^2c_\gamma^2
 =\tfrac12(1-2^{-\mathcal M_\gamma}),\qquad
 P_8(\gamma)=u_\gamma^2c_\gamma^2
 =1-2^{-\mathcal M_\gamma}.
 \label{eq:introConcentration}
\end{equation}
Thus the concentration probability is directly fixed by the nonlocal magic
profile, without introducing an additional communication-response quantity.
The resulting positive global stabilizer yield is consistent with vanishing
collectively optimized magic density because the operation classes differ.
Symmetry-resolved reduction \cite{SymmetryBH} further separates sector weights
from accessible within-sector entanglement; several copies activate the latter
under an explicitly added local parity restriction. A direct comparison of
two calibrated noisy preparations shows that equal target fidelity does not
determine the distillation rate. Gaussian code-distance constraints
\cite{FermionicQEC} distinguish the parity-protected right-state support from
a code capable of correcting arbitrary logical errors.

\rev{At the operator level, the same deformation can also be viewed as a
normalized metric-gradient trajectory generated by a complexified Majorana
bilinear, which separates the present nonunitary Gaussian map from a physical
fermionic Clifford gate \cite{Graefe2010,MagoulasEvangelista2026}. In the
state-sharing problem, the flat-spectrum construction of
Ref.~\cite{HuhtanenSwap} further gives an arbitrary-depth deterministic
alternative to the heralded identical-link recursion.}

\rev{The later many-copy comparisons use three further distinctions. The
stabilizer-fidelity definition of nonlocal magic is controlled by the
entanglement-energy capacity \cite{NonlocalMagicSpreading}, whereas
fixed-frame stabilizer entropy and the entropy-subtracted subsystem measure
support a separate entanglement-dominated/magic-dominated classification
\cite{MagicAreaLaws}. Total-parity twirling and local-unitary magic injection
\cite{MagicInjection} concern different free-operation classes; the formulas
below therefore specialize each construction separately rather than treating
them as interchangeable magic monotones.}

\section{Fixed bipartition and noisy right-state preparation}
\label{sec:model}
\subsection{Isospectral deformation in the fixed physical Hilbert space}
We fix the ordinary positive inner product on
\(\mathcal H_A\otimes\mathcal H_B=\mathbb C^2\otimes\mathbb C^2\), together with
the computational tensor-product basis. Let
\begin{equation}
 H_0=\diag(\lambda_0,\lambda_0,\lambda_1,\lambda_2),\quad
 S_\gamma=e^{\gamma X\otimes X},\quad
 H_\gamma=S_\gamma H_0S_\gamma^{-1},
 \label{eq:H}
\end{equation}
with distinct real \(\lambda_0,\lambda_1,\lambda_2\) and finite real \(\gamma\).
Since \(S_\gamma\) is invertible, the spectrum and the rank-two degeneracy of
\(\lambda_0\) are preserved; the Hamiltonian remains diagonalizable. For
\(\gamma\ne0\), the noncommuting deformation gives a non-Hermitian Hamiltonian.
There is no finite-\(\gamma\) exceptional point in this construction.

Write $h=\cosh\gamma$ and $k=\sinh\gamma$. Because
$(X\otimes X)^2=I$, the even and odd powers of the exponential give
\begin{equation}
 S_\gamma=hI+kX\otimes X,\qquad S_\gamma^{-1}=hI-kX\otimes X,
 \qquad \det S_\gamma=1.
 \label{eq:Sexplicit}
\end{equation}
Multiplying the three matrices in Eq.~\eqref{eq:H} yields
\begin{equation}
H_\gamma=\begin{pmatrix}
 h^2\lambda_0-k^2\lambda_2&0&0&hk(\lambda_2-\lambda_0)\\
 0&h^2\lambda_0-k^2\lambda_1&hk(\lambda_1-\lambda_0)&0\\
 0&hk(\lambda_0-\lambda_1)&h^2\lambda_1-k^2\lambda_0&0\\
 hk(\lambda_0-\lambda_2)&0&0&h^2\lambda_2-k^2\lambda_0
\end{pmatrix}.
\label{eq:Hexplicit}
\end{equation}
The off-diagonal entries are antisymmetric and real, displaying
non-Hermiticity without changing the characteristic polynomial:
\begin{equation}
 \det(zI-H_\gamma)=\det[S_\gamma(zI-H_0)S_\gamma^{-1}]
 =(z-\lambda_0)^2(z-\lambda_1)(z-\lambda_2).
 \label{eq:charpoly}
\end{equation}
For a computational reference vector $\ket n$, the right and left columns
are $\ket{R_n}=S_\gamma\ket n$ and $\ket{L_n}=S_\gamma^{-1}\ket n$.
Since $S_\gamma$ is Hermitian,
\begin{equation}
 H_\gamma\ket{R_n}=\lambda_n\ket{R_n},\quad
 H_\gamma^\dagger\ket{L_n}=\lambda_n\ket{L_n},\quad
 \langle L_n|R_j\rangle=\delta_{nj},\quad
 \sum_n\ket{R_n}\bra{L_n}=I.
 \label{eq:biorthbasis}
\end{equation}
Here $\lambda_{00}=\lambda_{01}=\lambda_0$, $\lambda_{10}=\lambda_1$,
and $\lambda_{11}=\lambda_2$. Explicitly,
\begin{align}
 R_{00}&=h\ket{00}+k\ket{11},&R_{01}&=h\ket{01}+k\ket{10},\notag\\
 R_{10}&=h\ket{10}+k\ket{01},&R_{11}&=h\ket{11}+k\ket{00}.
 \label{eq:allright}
\end{align}
Each right vector has squared norm $h^2+k^2=\cosh2\gamma$; the left
vectors follow by replacing $k$ by $-k$. The eigenvalues of $S_\gamma$
are $e^{\gamma}$ and $e^{-\gamma}$, each twice, so its spectral condition
number is $e^{2|\gamma|}$. Large deformation is an ill-conditioned limit,
not a finite-parameter coalescence of eigenvectors.

Transporting the fixed reference vector \(\ket{00}\), then normalizing in the
ordinary inner product, gives
\begin{equation}
 \ket{\psi_\gamma}=
 \frac{\cosh\gamma\ket{00}+\sinh\gamma\ket{11}}{\sqrt{\cosh2\gamma}}.
 \label{eq:psi}
\end{equation}
A local phase removes the sign of \(\gamma\), so our state diagnostics use
\begin{equation}
 u_\gamma=\sech(2|\gamma|),\qquad
 c_\gamma=\tanh(2|\gamma|),\qquad
 p_\pm=\frac{1\pm u_\gamma}{2},\quad u_\gamma^2+c_\gamma^2=1.
 \label{eq:uc}
\end{equation}
\rev{The normalized deformation path has a useful geometric interpretation.
For \(G=X\otimes X\) and the signed-\(\gamma\) state in
Eq.~\eqref{eq:psi}, differentiation of the normalized exponential gives}
\begin{equation}
 \rev{\frac{\partial}{\partial\gamma}\ket{\psi_\gamma}
 =\bigl(G-\langle G\rangle_\gamma\bigr)\ket{\psi_\gamma}.}
 \label{eq:deformationGradient}
\end{equation}
\rev{Equation~\eqref{eq:deformationGradient} is the pure metric-gradient part
of normalized non-Hermitian projective dynamics, analogous to the gradient
flow appearing in non-Hermitian dimer descriptions
\cite{Graefe2010}. On the logical Bloch circle of
\(\operatorname{span}\{\ket{00},\ket{11}\}\), put
\(x_\gamma=\langle G\rangle_\gamma=\tanh(2\gamma)\) and
\(z_\gamma=\langle Z_A\rangle_\gamma=\sech(2\gamma)\). Then}
\begin{equation}
 \rev{\frac{d x_\gamma}{d\gamma}=2(1-x_\gamma^2)=2z_\gamma^2,
 \qquad
 \frac{d z_\gamma}{d\gamma}=-2x_\gamma z_\gamma.}
 \label{eq:logicalGradientFlow}
\end{equation}
\rev{Here \(\gamma\) remains a deformation coordinate rather than physical
time; no Bose--Hubbard mean-field limit or semiclassical correspondence is
assumed.}
This selects a canonical transported eigenvector, not a basis-independent
resource of the whole degenerate eigenspace. Other linear combinations in
that eigenspace can have different entanglement. Also, the positive metric
\(S_\gamma^{-2}\) that makes \(H_\gamma\) quasi-Hermitian is not used to redefine
the physical inner product or measurement probabilities here. The operational
state is the positive normalized right-state density matrix
\(\ket{\psi_\gamma}\bra{\psi_\gamma}\), not a biorthogonal reduction.

The distinction between the two reductions can be checked directly.
For the canonical reference vector,
\begin{equation}
 \Tr_B(\ket{R_{00}}\bra{L_{00}})
 =\diag(h^2,-k^2),\qquad
 \Tr_B(\ket{\psi_\gamma}\bra{\psi_\gamma})=\diag(p_+,p_-).
 \label{eq:tworeductions}
\end{equation}
The first operator has trace one but a negative eigenvalue when $\gamma\ne0$;
it is not a Born-rule density matrix. The second is positive and supplies all
probabilities and resource measures below. The positive metric
$G_\gamma=S_\gamma^{-2}$ satisfies
$H_\gamma^\dagger G_\gamma=G_\gamma H_\gamma$ by matrix multiplication,
but changing to that metric would change the operational question.

The doublet also makes the eigenvector-selection issue explicit. Put
$a=h/\sqrt{h^2+k^2}$, $b=k/\sqrt{h^2+k^2}$, and
$\ket{\psi_1}=a\ket{01}+b\ket{10}$. The normalized doublet state
$\alpha\ket{\psi_\gamma}+\beta\ket{\psi_1}$ has coefficient matrix
\begin{equation}
 C_{\alpha\beta}=\begin{pmatrix}a\alpha&a\beta\\b\beta&b\alpha\end{pmatrix},
 \quad |\alpha|^2+|\beta|^2=1,\quad
 C(\alpha\psi_\gamma+\beta\psi_1)=2|ab|\,|\alpha^2-\beta^2|.
 \label{eq:doubletconcurrence}
\end{equation}
For nonzero deformation, the combinations $\alpha=\pm\beta$ are product
states in the same degenerate eigenspace. Thus the canonical transport,
the fixed bipartition, and the chosen positive inner product are essential
parts of the model definition.

For comparison, the ordinary singular values of the canonical biorthogonal
reduction in Eq.~\eqref{eq:tworeductions} are $h^2,k^2$. Their normalization
gives the original Schmidt weights \cite{Wu}:
\begin{equation}
 \frac{(h^2,k^2)}{h^2+k^2}=(p_+,p_-),\qquad
 S_{\mathrm{SVD}}=h_2(p_+).
 \label{eq:inheritedSVD}
\end{equation}
If independent diagonal unit balancing is used instead, then for
$\gamma\ne0$, $D=\operatorname{diag}(h^{-1},|k|^{-1})$ gives
$D\operatorname{diag}(h^2,-k^2)D=\operatorname{diag}(1,-1)$.
The balanced singular-value entropy is therefore one bit at every nonzero
deformation and zero at the rank-one product endpoint. It is a
rank-sensitive invariant under the stated rescalings, not the continuous
physical right-state entanglement. These complementary reductions retain
their distinct definitions; the Born distribution below uses only the
positive right-state density matrix.

\subsection{A two-parameter operational extension}
Consider the real maximally correlated state
\begin{equation}
 \rho(u,t)=\frac12
 \begin{pmatrix}
 1+u&0&0&t\\0&0&0&0\\0&0&0&0\\t&0&0&1-u
 \end{pmatrix},\qquad
 0\le u\le1,\quad0\le t\le c:=\sqrt{1-u^2}.
 \label{eq:rho}
\end{equation}
The original pure state is the boundary \((u,t)=(u_\gamma,c_\gamma)\).
Applying phase noise with visibility \(\eta\in[0,1]\) to either qubit gives
\begin{equation}
 \rho_{\gamma,\eta}=\rho(u_\gamma,\eta c_\gamma).
 \label{eq:dephasing}
\end{equation}
Considering subject subsystem $B$ of the pure right eigenstate $\rho(u_\gamma, c_\gamma)$ to a local phase-flip channel $\mathcal{E}_p^{(B)}(\rho) = (1-p)\rho + p (\mathbb{I}_A \otimes Z_B)\rho (\mathbb{I}_A \otimes Z_B)$, where $p \in [0, 1/2]$ represents the single-qubit phase-flip probability. In the ordered computational product basis $\{\vert{}00\rangle, \vert{}01\rangle, \vert{}10\rangle, \vert{}11\rangle\}$, the operator $\mathbb{I}_A \otimes Z_B = \operatorname{diag}(1, -1, 1, -1)$ acts on the diagonal populations since $Z_B \vert{}00\rangle = \vert{}00\rangle$ and $Z_B \vert{}11\rangle = -\vert{}11\rangle$, yielding $(\mathbb{I}_A \otimes Z_B)\vert{}00\rangle\langle 00\vert{}(\mathbb{I}_A \otimes Z_B) = \vert{}00\rangle\langle 00\vert{}$ and $(\mathbb{I}_A \otimes Z_B)\vert{}11\rangle\langle 11\vert{}(\mathbb{I}_A \otimes Z_B) = \vert{}11\rangle\langle 11\vert{}$. For off-diagonal coherences, $(\mathbb{I}_A \otimes Z_B)\vert{}00\rangle\langle 11\vert{}(\mathbb{I}_A \otimes Z_B) = -\vert{}00\rangle\langle 11\vert{}$,
$(\mathbb{I}_A \otimes Z_B)\vert{}11\rangle\langle 00\vert{}(\mathbb{I}_A \otimes Z_B) = -\vert{}11\rangle\langle 00\vert{}$.
Thus the density matrix evolves into
$$\mathcal{E}_p^{(B)}[\rho(u_\gamma, c_\gamma)] = \frac{1}{2} \begin{pmatrix} 1+u_\gamma & 0 & 0 & (1-2p)c_\gamma \\ 0 & 0 & 0 & 0 \\ 0 & 0 & 0 & 0 \\ (1-2p)c_\gamma & 0 & 0 & 1-u_\gamma \end{pmatrix} = \rho(u_\gamma, (1-2p)c_\gamma)$$.
Thus the off-diagonal coherence $(1-2p)c_\gamma$ has visibility parameter $\eta c_\gamma$ and the effective probability of applying the phase kick is $p =p_-= \frac{1-\eta}{2}$.
At the complete dephasing boundary $p = 1/2$ ($\eta = 0$ and $t = 0$), the surviving coherence factor vanishes, causing all off-diagonal matrix elements to collapse. The resulting state reduces to   
\begin{equation}
\rho(u_\gamma, 0) = \frac{1+u_\gamma}{2}\vert{}00\rangle\langle 00\vert{} + \frac{1-u_\gamma}{2}\vert{}11\rangle\langle 11\vert{} = p_+ \vert{}0\rangle_A\langle 0\vert{}_A \otimes \vert{}0\rangle_B\langle 0\vert{}_B + p_- \vert{}1\rangle_A\langle 1\vert{}_A \otimes \vert{}1\rangle_B\langle 1\vert{}_B 
\end{equation}
which is a classical mixture of orthogonal product states. Since $\rho(u_\gamma, 0)$ is diagonal in the computational product basis, it lacks any non-classical superposition across the bipartition. Evaluating its partial transpose yields purely non-negative eigenvalues $\{p_+, p_-, 0, 0\}$, confirming that the PPT negativity vanishes identically, $\mathcal{N}(\rho) = t/2 = 0$. Furthermore, its distillable entanglement satisfies $\mathcal{D}(u_\gamma, 0) = h_2(p_+) - h_2(p_+) = 0$. Thus, at $p = 1/2$, the non-unitary doublet completely exhausts its quantum entanglement resources and degrades into a classically correlated separable state.
Here the probability of applying \(Z\) is \((1-\eta)/2\). This is preparation
noise acting on an eigenstate; the mixed state is not asserted to be a right
eigenvector, and the noisy preparation does not alter Eq.~\eqref{eq:H}.
The spectrum of \(\rho(u,t)\) is
\begin{equation}
 \left\{\frac{1+s}{2},\frac{1-s}{2},0,0\right\},\quad s=\sqrt{u^2+t^2},
 \qquad S(\rho_B)=\hbin(p_+).
 \label{eq:rhospectrum}
\end{equation}
Partial transposition has eigenvalues \(p_+,p_-,t/2,-t/2\), hence
\begin{equation}
 \mathcal N(\rho)=\frac t2,\qquad C(\rho)=t.
 \label{eq:neg}
\end{equation}
Every state with \(t>0\) is NPT; \(t=0\) is a separable, potentially classically
correlated state. This two-qubit family contains no PPT bound entangled state.

The support block of Eq.~\eqref{eq:rho} is
$\frac12(I+uZ+tX)$ in the ordered even-parity basis $(\ket{00},\ket{11})$.
Its determinant is $(1-u^2-t^2)/4$, which gives the physical disk directly.
In the full qubit Pauli basis,
\begin{equation}
 \rho(u,t)=\frac14[I\otimes I+u(Z\otimes I+I\otimes Z)
                  +Z\otimes Z+t(X\otimes X-Y\otimes Y)].
 \label{eq:rhoPauli}
\end{equation}
To obtain this Pauli-basis expansion, we write the two-parameter density matrix of Eq. (\ref{eq:rho}) in the computational basis as $$\rho(u,t) = p_{+}\vert{}00\rangle\langle 00\vert{} + p_{-}\vert{}11\rangle\langle 11\vert{} + \frac{t}{2}\big(\vert{}00\rangle\langle 11\vert{} + \vert{}11\rangle\langle 00\vert{}\big), \qquad p_{\pm} = \frac{1\pm u}{2}.$$
Expressing the single-qubit computational projectors as $\vert{}0\rangle\langle 0\vert{} = (I+Z)/2$ and $\vert{}1\rangle\langle 1\vert{} = (I-Z)/2$ yields the diagonal population operators
\begin{equation}
\begin{aligned}
 \vert{}00\rangle\langle 00\vert{} &= \frac{I+Z}{2}\otimes\frac{I+Z}{2} = \frac{1}{4}(I\otimes I + Z\otimes I + I\otimes Z + Z\otimes Z), \\  \vert{}11\rangle\langle 11\vert{} &= \frac{I-Z}{2}\otimes\frac{I-Z}{2} = \frac{1}{4}(I\otimes I - Z\otimes I - I\otimes Z + Z\otimes Z).  \end{aligned}\end{equation}
Similarly, using $X = \vert{}0\rangle\langle 1\vert{} + \vert{}1\rangle\langle 0\vert{}$ and $Y = -i\vert{}0\rangle\langle 1\vert{} + i\vert{}1\rangle\langle 0\vert{}$, we have
\begin{equation}
\begin{aligned}  
X\otimes X &= \vert{}00\rangle\langle 11\vert{} + \vert{}11\rangle\langle 00\vert{} + \vert{}01\rangle\langle 10\vert{} + \vert{}10\rangle\langle 01\vert{}, \\  Y\otimes Y &= -\vert{}00\rangle\langle 11\vert{} - \vert{}11\rangle\langle 00\vert{} + \vert{}01\rangle\langle 10\vert{} + \vert{}10\rangle\langle 01\vert{},  
\end{aligned}
\end{equation}
so subtracting the two eliminates the odd-parity terms and gives $\vert{}00\rangle\langle 11\vert{} + \vert{}11\rangle\langle 00\vert{} = \frac{1}{2}(X\otimes X - Y\otimes Y)$. Substituting these three operator identities back into $\rho(u,t)$ and using $p_{+} + p_{-} = 1$ together with $p_{+} - p_{-} = u$ directly produces Eq. (\ref{eq:rhoPauli}).
On the pure-state boundary $(u,t) = (u_{\gamma}, c_{\gamma})$ with $u_{\gamma}^{2} + c_{\gamma}^{2} = 1$, this expansion factorizes into a product of two commuting generalized-stabilizer projectors. Specifically, define the two Hermitian operators
$$S_{1} = Z\otimes Z, \qquad S_{2} = u_{\gamma}(Z\otimes I) + c_{\gamma}(X\otimes X).$$
They satisfy every requirement of a complete set of commuting involutory stabilizers on $\mathbb{C}^{2}\otimes\mathbb{C}^{2}$:Involution ($S_{i}^{2} = I\otimes I$): Clearly $S_{1}^{2} = Z^{2}\otimes Z^{2} = I\otimes I$. Because $\{Z, X\} = 0$, the anticommutator $\{Z\otimes I, X\otimes X\} = \{Z, X\}\otimes X = 0$ vanishes, so $u_{\gamma}^{2} + c_{\gamma}^{2} = 1$ gives $$S_{2}^{2} = u_{\gamma}^{2}(Z^{2}\otimes I) + c_{\gamma}^{2}(X^{2}\otimes X^{2}) + u_{\gamma}c_{\gamma}\{Z\otimes I, X\otimes X\} = (u_{\gamma}^{2} + c_{\gamma}^{2})(I\otimes I) = I\otimes I.$$
Both $S_{1}$ and $S_{2}$ are therefore Hermitian involutions with eigenvalues $\pm 1$. Mutual commutativity ($[S_{1}, S_{2}] = 0$): While $[Z\otimes Z, Z\otimes I] = 0$ trivially, $(Z\otimes Z)(X\otimes X) = (ZX)\otimes(ZX) = (-iY)\otimes(-iY) = -Y\otimes Y = (XZ)\otimes(XZ) = (X\otimes X)(Z\otimes Z)$, confirming $[S_{1}, S_{2}] = 0$.
Independence and state stabilization: Since $\operatorname{Tr}(S_{1}) = \operatorname{Tr}(S_{2}) = \operatorname{Tr}(S_{1}S_{2}) = 0$, the operators $S_{1}$ and $S_{2}$ are algebraically independent and uniquely select a simultaneous $+1$ eigenstate. In the positive-phase convention $\vert{}\psi_{\gamma}\rangle = \sqrt{p_{+}}\vert{}00\rangle + \sqrt{p_{-}}\vert{}11\rangle$ (where $p_{+} - p_{-} = u_{\gamma}$ and $2\sqrt{p_{+}p_{-}} = c_{\gamma}$), one verifies directly that $S_{1}\vert{}\psi_{\gamma}\rangle = \vert{}\psi_{\gamma}\rangle$ and $S_{2}\vert{}\psi_{\gamma}\rangle = (u_{\gamma}\sqrt{p_{+}} + c_{\gamma}\sqrt{p_{-}})\vert{}00\rangle + (-u_{\gamma}\sqrt{p_{-}} + c_{\gamma}\sqrt{p_{+}})\vert{}11\rangle = \vert{}\psi_{\gamma}\rangle$.
Multiplying the two commuting projectors $(I + S_{1})/2$ and $(I + S_{2})/2$ and using $(Z\otimes Z)(Z\otimes I) = I\otimes Z$ together with $(Z\otimes Z)(X\otimes X) = -Y\otimes Y$ yields
\begin{equation}
\begin{aligned}  
\frac{I + S_{1}}{2}\frac{I + S_{2}}{2} &= \frac{1}{4}\Big[I\otimes I + Z\otimes Z + u_{\gamma}(Z\otimes I) + u_{\gamma}(Z\otimes Z)(Z\otimes I) \\ &\qquad + c_{\gamma}(X\otimes X) + c_{\gamma}(Z\otimes Z)(X\otimes X)\Big] \\  &= \frac{1}{4}\big[I\otimes I + u_{\gamma}(Z\otimes I + I\otimes Z) + Z\otimes Z + c_{\gamma}(X\otimes X - Y\otimes Y)\big] \\  &= \rho(u_{\gamma}, c_{\gamma}) = \vert{}\psi_{\gamma}\rangle\langle\psi_{\gamma}\vert{}. 
\end{aligned}
\end{equation}
In the noisy interior $t < c_{\gamma}$, replacing $c_{\gamma}$ by $t$ in $S_{2}$ gives $S_{2}(u,t)^{2} = (u^{2}+t^{2})(I\otimes I) < I\otimes I$, reflecting the loss of purity ($\operatorname{Tr}\rho^{2} = (1+u^{2}+t^{2})/2 < 1$) as the rank-one pure projector dephases into a rank-two mixed state.

The characteristic polynomial of the support block is
$z^2-z+(1-u^2-t^2)/4$, giving Eq.~\eqref{eq:rhospectrum} and
$\Tr\rho^2=(1+u^2+t^2)/2$. Partial transpose moves the two off-diagonal
entries to the $(\ket{01},\ket{10})$ block:
\begin{equation}
 \rho^{\Gamma_B}=\begin{pmatrix}p_+&0&0&0\\0&0&t/2&0\\
 0&t/2&0&0\\0&0&0&p_-\end{pmatrix},\qquad
 \|\rho^{\Gamma_B}\|_1=1+t.
 \label{eq:PTexplicit}
\end{equation}
This proves the negativity in Eq.~\eqref{eq:neg}; the two-qubit
$X$-state concurrence is $2\max(0,|\rho_{00,11}|-
\sqrt{\rho_{01,01}\rho_{10,10}})=t$. For the pure boundary the Schmidt
coefficients are $\sqrt{p_+},\sqrt{p_-}$, hence $2\sqrt{p_+p_-}=c$.
For $Z_B = I \otimes Z=\operatorname{diag}(1, -1, 1, -1)$, $Z_B \vert{}00\rangle = \vert{}00\rangle$ and $Z_B \vert{}11\rangle = -\vert{}11\rangle$, $$Z_B \rho(u, t) Z_B = \frac{1}{2}\begin{pmatrix} 1+u & 0 & 0 & -t \\ 0 & 0 & 0 & 0 \\ 0 & 0 & 0 & 0 \\ -t & 0 & 0 & 1-u \end{pmatrix}$$, and the channel reads$$\mathcal{E}(\rho) = (1-p)\rho(u,t) + p Z_B \rho(u,t) Z_B= \frac{1}{2}\begin{pmatrix} 1+u & 0 & 0 & (1-2p)t \\ 0 & 0 & 0 & 0 \\ 0 & 0 & 0 & 0 \\ (1-2p)t & 0 & 0 & 1-u \end{pmatrix}$$.
It give
\begin{equation}
 (1-p)\rho(u,c)+pZ_B\rho(u,c)Z_B=\rho(u,(1-2p)c),
 \quad 0\le p\le\tfrac12.
 \label{eq:localchannelexplicit}
\end{equation}
The noise parameter is a phase-flip probability, so increasing it beyond
$1/2$ would restore coherence with a local sign change rather than produce
monotonic further dephasing.

\paragraph{Optimized CHSH nonlocality.}
The local Bloch vectors of \(\rho(u,t)\) are \((0,0,u)\), and its correlation
matrix is \(T=\operatorname{diag}(t,-t,1)\). The Horodecki criterion therefore
gives \cite{Horodecki}
\begin{equation}
 \mathcal B_{\max}(u,t)=2\sqrt{1+t^2}.
 \label{eq:chsh}
\end{equation}
It is attained by
\[
 A_0=Z,\quad A_1=X,\qquad
 B_0=\frac{Z+tX}{\sqrt{1+t^2}},\quad
 B_1=\frac{Z-tX}{\sqrt{1+t^2}}.
\]
Consequently every state in this idealized family with \(t>0\) violates
CHSH, whereas the completely dephased boundary \(t=0\) is both separable and
CHSH classical. This optimized Bell value is a property of the state under
local measurement optimization; no auxiliary communication protocol is used.

\subsection{Correlated parity noise and a finite extension}
\label{sec:paritynoise}
Pairwise phase noise \cite{FechisinPower} composes
maps $\mathcal E_{ij;q}(\rho)=(1-q)\rho+qZ_iZ_j\rho Z_iZ_j$ over many
spatially separated pairs. Its distance-dependent probabilities and
thermodynamic limit are part of its many-body setting. For the present
state, let $\Pi_+=(I+Z_AZ_B)/2$. Since
$\rho(u,t)=\Pi_+\rho(u,t)\Pi_+$, one obtains the exact finite-system result
\begin{equation}
 \mathcal E_{AB;q}[\rho(u,t)]=\rho(u,t)\quad\text{for every }q\in[0,1].
 \label{eq:parityimmune}
\end{equation}
The support carries one eigenvalue of the noise operator, so the kick is
a scalar there. The mechanism is operator algebra, not a claim that
two-qubit states are immune to all correlated noise. In fact any state
commuting with $Z_AZ_B$ is fixed, including mixtures of parity sectors.

Both $Z_AZ_B\ket{00}=\ket{00}$ and $Z_AZ_B\ket{11}=\ket{11}$ hold,
so the phase kick has no relative phase within this support. In particular
the immunity holds after local dephasing as well as before it. With
$\Pi_\pm=(I\pm Z_AZ_B)/2$ and $\rho_{rs}=\Pi_r\rho\Pi_s$,
\begin{equation}
 \mathcal E_{AB;q}(\rho)=\rho_{++}+\rho_{--}
                    +(1-2q)(\rho_{+-}+\rho_{-+}).
 \label{eq:parityblocks}
\end{equation}
Only coherence between opposite parity sectors is attenuated. Definite
parity is sufficient for protection; mixtures without inter-sector coherence
are protected too. This finite-dimensional fixed-point algebra is distinct
from strong-to-weak spontaneous symmetry breaking in a spatial thermodynamic
limit \cite{FechisinPower}. For the two specified channels, the loss of
entanglement comes from $Z_B$ noise while the $Z_AZ_B$ channel leaves it
unchanged. This statement is about those channels, rather than a claim that
all entanglement degradation must originate in local noise.

More generally, a Pauli phase channel with probabilities
$w_0,w_A,w_B,w_{AB}$ for $I,Z_A,Z_B,Z_AZ_B$ preserves the populations
and multiplies the coherence by
\begin{equation}
 \eta_{\mathrm{eff}}=w_0+w_{AB}-w_A-w_B=1-2(w_A+w_B).
 \label{eq:phaseweights}
\end{equation}
A negative multiplier is removed by the same local phase convention, giving
$t'=|\eta_{\mathrm{eff}}|t$. A single local kick suppresses coherence; a
joint kick does not. Within this family CHSH exceeds two for every $t'>0$,
and separability is reached only when $t'=0$. These are neither SW-SSB
transitions nor many-body phase boundaries; they are finite-state statements
about coherence, Bell violation, and separability.

The immunity is sensitive to coherent parity leakage. With
$\ell=\Tr(\Pi_-\widetilde\rho)$ and $C=\Pi_+\widetilde\rho\Pi_-$, block
decomposition and positivity give
\begin{equation}
 \|\mathcal E_{AB;q}(\widetilde\rho)-\widetilde\rho\|_1
 =4q\|C\|_1\le4q\sqrt{\ell(1-\ell)}.
 \label{eq:leakagebound}
\end{equation}
Incoherent leakage with $C=0$ is still fixed; leakage population alone
does not determine the noise effect.

To obtain Eq.~\eqref{eq:leakagebound}, write the density matrix as
$\left(\begin{smallmatrix}A&C\\C^\dagger&B\end{smallmatrix}\right)$
in its parity decomposition. The difference after the channel is
$-2q\left(\begin{smallmatrix}0&C\\C^\dagger&0\end{smallmatrix}\right)$;
the singular values of $C$ occur twice in this Hermitian block. Positivity
allows $C=A^{1/2}KB^{1/2}$ with $\|K\|_\infty\le1$, and the Schatten
Cauchy--Schwarz inequality gives
$\|C\|_1\le\|A^{1/2}\|_2\|B^{1/2}\|_2
=\sqrt{(1-\ell)\ell}$. No population-only approximation enters this bound.

There is a nontrivial finite enlargement without importing a many-body
phase diagram. Take $\ket\Psi=\ket{\psi_\gamma}_{12}\ket{\psi_\gamma}_{34}$.
The within-dimer kick $Z_1Z_2$ is scalar on its support, whereas the
cross-dimer kick $Z_1Z_3$ has expectation $u^2$. Applying just that pair
channel gives
\begin{equation}
 \Tr\left[\mathcal E_{13;q}(\ket\Psi\bra\Psi)^2\right]
 =1-2q(1-q)(1-u^4).
 \label{eq:crossdimerpurity}
\end{equation}
Thus cross-pair noise can mix a product of protected dimers even while
their total parity stays definite. Conversely extended GHZ states are
fixed by all the pair kicks, since every $Z_iZ_j$ is scalar on their support. A spatial bulk is needed for its phase diagram, but
spatial extension alone does not guarantee a nontrivial noise effect.

For the cross-dimer test, let $U=Z_1Z_3$ and
$\omega=\ket\Psi\bra\Psi$. Since $\omega^2=\omega$,
\begin{align}
 \Tr[( (1-q)\omega+qU\omega U)^2]
 &=(1-q)^2+q^2+2q(1-q)|\langle\Psi|U|\Psi\rangle|^2,\notag\\
 \langle\Psi|U|\Psi\rangle
 &=\langle\psi_\gamma|Z_1|\psi_\gamma\rangle
   \langle\psi_\gamma|Z_3|\psi_\gamma\rangle=u^2.
 \label{eq:crossdimerproof}
\end{align}
Thus the overlap of the kicked and un-kicked pure states is $u^4$, proving
Eq.~\eqref{eq:crossdimerpurity}.

\section{Pure-state nonlocal magic profile}
\label{sec:puremagic}
For a pure two-qubit state define the stabilizer R\'enyi entropy in the fixed
Pauli frame by
\begin{equation}
 M_2^S(\psi)=-\log_2\zeta_4(\psi),\qquad
 \zeta_4(\psi)=\frac14\sum_{P\in\{I,X,Y,Z\}^{\otimes2}}
                      \langle P\rangle_\psi^4.
 \label{eq:SRE}
\end{equation}
Its nonlocal version minimizes over unrestricted local unitaries. For the
present rank-two pure family, the Schmidt representative is optimal
\cite{Wu,Sierant}. Its nonzero Pauli expectations are
\(1,u,u,1,c,-c\), and hence
\begin{equation}
 \Mcal_\gamma=M_2^{\NL}(\psi_\gamma)=M_2^S(\psi_\gamma)
 =-\log_2(1-u_\gamma^2+u_\gamma^4).
 \label{eq:magic}
\end{equation}
The fixed Schmidt frame therefore attains the local-unitary minimum. The
profile is maximal when
\begin{equation}
 u_\gamma^2=c_\gamma^2=\frac12,\qquad
 |\gamma_*|=\frac12\operatorname{arcosh}\sqrt2,\qquad
 \Mcal_{\gamma_*}=\log_2\frac43.
 \label{eq:extrema}
\end{equation}

\begin{figure}[htbp]
\centering
\includegraphics[width=0.75\linewidth]{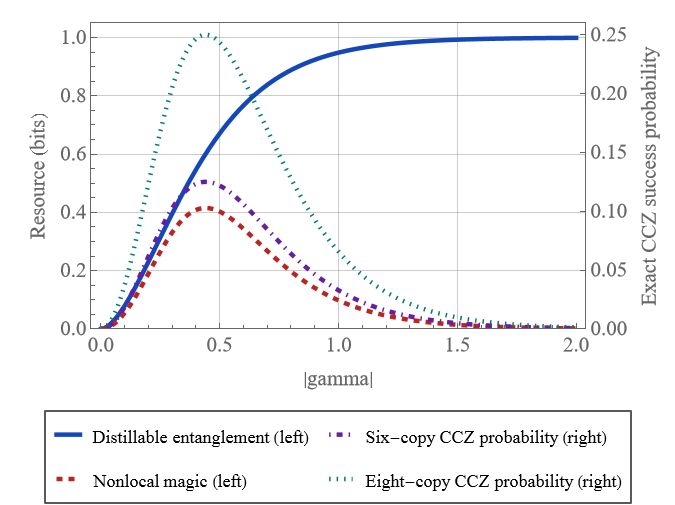}
\caption{Pure right-state resources and exact universal concentration.
The left axis gives $E_d(\psi_\gamma)$ and $\mathcal M_\gamma$ in bits;
the right axis gives the six- and eight-copy CCZ success probabilities.
The probabilities are calculated from Eq.~\eqref{eq:CCZsix} and
Eq.~\eqref{eq:CCZeight}, and peak at the deformation maximizing magic.
Their direct relation to nonlocal magic is
Eq.~\eqref{eq:magicconcentrationbridge}. The concentration operation class
is global qubit stabilizer processing.}
\label{fig:pure}
\end{figure}

Figure~\ref{fig:pure} compares this nonlocal magic with the pure-state
distillable entanglement and the exact concentration probabilities derived in
Sec.~\ref{sec:magicconcentration}. The latter obey the direct identities in
Eq.~\eqref{eq:magicconcentrationbridge}; no communication protocol enters
that relation.

\section{Exact distillation and PPT regularizations}
\label{sec:distillation}
\subsection{Hashing meets a relative-entropy comparison}
Let \(E_d^\to,E_d,E_d^{\PPT}\) denote the asymptotic rates of Bell-pair
distillation under one-way LOCC, unrestricted LOCC, and completely
PPT-preserving maps. For clarity, define the ordinary PPT relative entropy and
the Rains comparison set by
\begin{align}
 D_{\PPT}(\rho)&=\inf_{\sigma\ge0,\,\Tr\sigma=1,\,\sigma^{\Gamma_B}\ge0}
                              D(\rho\|\sigma),\label{eq:PPTD}\\
 \PPT_1&=\{\tau\ge0:\|\tau^{\Gamma_B}\|_1\le1\},\qquad
 D_{\PPT_k}(\rho)=\inf_{\tau\in\PPT_k}D(\rho\|\tau).
 \label{eq:Rains}
\end{align}
Here \(D(\rho\|\tau)=\Tr\rho(\log_2\rho-\log_2\tau)\), including
subnormalized \(\tau\). For \(k\ge2\), the PPT comparison hierarchy \cite{Ao2026} is
defined recursively by requiring some \(\eta\in\PPT_{k-1}\) with
\(-\eta\le\tau^{\Gamma_B}\le\eta\). It satisfies
\begin{equation}
 \PPT\subseteq\PPT_{k+1}\subseteq\PPT_k\subseteq\PPT_1.
 \label{eq:hierarchy}
\end{equation}

\begin{theorem}[Exact rate for the noisy deformed family]
For every physical \(\rho(u,t)\) in Eq.~\eqref{eq:rho},
\begin{equation}
 E_d^\to(\rho)=E_d(\rho)=E_d^{\PPT}(\rho)=\Dcal(u,t),
 \label{eq:exactEd}
\end{equation}
with \(\Dcal\) given by Eq.~\eqref{eq:introD}. For every integer \(n\ge1\) and
every finite hierarchy level \(k\ge1\),
\begin{equation}
 D_{\PPT_k}(\rho^{\otimes n})=D_{\PPT}(\rho^{\otimes n})=n\Dcal(u,t).
 \label{eq:exactDn}
\end{equation}
Consequently all these regularized relative entropies equal \(\Dcal(u,t)\).
\end{theorem}
\begin{proof}
The hashing inequality \cite{DevetakWinter} gives
\begin{equation}
 E_d^\to(\rho)\ge I_c(A|B)_\rho
 =S(\rho_B)-S(\rho)=\Dcal(u,t).
 \label{eq:hashing}
\end{equation}
The diagonal state
\begin{equation}
 \sigma(u)=p_+\ket{00}\bra{00}+p_-\ket{11}\bra{11}
 \label{eq:sigma}
\end{equation}
is separable and hence belongs to every comparison set in
Eq.~\eqref{eq:hierarchy}. Direct evaluation gives
\begin{equation}
 D(\rho\|\sigma)=\hbin(p_+)-S(\rho)=\Dcal(u,t).
 \label{eq:Dcomparison}
\end{equation}
In the support of $\rho$, $\log_2\sigma$ is diagonal with entries
$\log_2p_+$ and $\log_2p_-$. The off-diagonal entries of $\rho$ do not
contribute to its trace against that diagonal operator:
\begin{equation}
 -\Tr\rho\log_2\sigma=-p_+\log_2p_+-p_-\log_2p_-=h_2(p_+),
 \quad\Tr\rho\log_2\rho=-h_2((1+s)/2).
 \label{eq:comparisontrace}
\end{equation}
Zero-weight terms are taken by continuity. The comparison has the same
diagonal populations as the original state, so its relative entropy removes
exactly the support-block entropy while retaining the marginal entropy.

The Rains relative entropy upper-bounds PPT distillation \cite{Rains}.
Thus
\begin{equation}
 \Dcal\le E_d^\to\le E_d\le E_d^{\PPT}
 \le D_{\PPT_1}\le D_{\PPT_k}\le D_{\PPT}\le\Dcal.
 \label{eq:sandwich}
\end{equation}
All quantities are equal. For \(\rho^{\otimes n}\), the product comparison
state \(\sigma^{\otimes n}\) gives the upper bound \(n\Dcal\), while hashing
gives the lower bound \(S(B^n)-S(A^nB^n)=n\Dcal\). The same sandwich proves
Eq.~\eqref{eq:exactDn}. Endpoint values follow by continuity, or directly
for the pure product state. This is the maximally correlated-state mechanism
specialized to the model, not a claim that hashing is optimal for all states.
\end{proof}

For the original deformation with phase visibility \(\eta\), the explicit rate is
\begin{equation}
 \Dcal_{\gamma,\eta}=
 \hbin\!\left(\frac{1+\sech2|\gamma|}{2}\right)
 -\hbin\!\left(\frac{1+
 \sqrt{\sech^2(2|\gamma|)+\eta^2\tanh^2(2|\gamma|)}}2\right).
 \label{eq:Dgammaeta}
\end{equation}
It is strictly positive for \(\gamma\ne0\) and \(\eta>0\). At \(\eta=1\) it
reduces to the pure entanglement entropy. At complete dephasing it vanishes,
and at the Bell limit it tends to \(1-\hbin((1+\eta)/2)\).

\subsection{Unknown-state universal attainment}
Universal unknown-state LOCC distillation \cite{Takagi2026} attains
the calculated rate, without assuming that $\gamma$ or the noisy state is
known beforehand. For unknown iid input $\rho_{\gamma,\eta}$, its
state-independent protocol asymptotically attains $\mathcal D_{\gamma,\eta}$.
On the pure boundary,
\begin{equation}
 R_{\mathrm{univ}}(\gamma)=h_2\!\left(\frac{1-\sech2|\gamma|}{2}\right)
 \ge1-\sech2|\gamma|=P_{\mathrm{filter}}(\gamma).
 \label{eq:universalPure}
\end{equation}
Here the probability is the Schmidt weight $p_-$, rather than the
phase-flip probability $p$ of Eq.~\eqref{eq:dephasing}. Concavity gives
\begin{equation}
 h_2(p_-)\ge(1-2p_-)h_2(0)+2p_-h_2(1/2)=2p_-,
 \qquad 0\le p_-\le\tfrac12.
 \label{eq:SchmidtChordDetailed}
\end{equation}
It is strict for $0<p_-<1/2$, and $2p_-=1-u_\gamma$ is the filter
success probability, proving Eq.~\eqref{eq:universalPure}. Universal attainment is an
asymptotic existence statement, not a deterministic finite-copy yield or
a computational-efficiency guarantee. The noisy filtering comparison and
the finite-sample model certificate remain different tasks.

\subsection{Correlated catalysts do not improve this rate}
In correlated catalytic manipulation the initial catalyst is independent of
the system, its marginal is returned exactly, and the final catalyst may be
correlated with the output. In the approximation convention for correlated PPT catalysis
\cite{Ao2026}, the output system can approach its target in trace
distance, while the catalyst marginal remains unchanged. The allowed maps are
completely PPT-preserving across the Alice--Bob cut including the catalyst.
Let \(E_d^{\PPT,\cc}\) denote the associated asymptotic Bell-pair rate.

\begin{corollary}[No catalytic gain in Bell-pair distillation]
Under the preceding conventions,
\begin{equation}
 E_d^{\PPT,\cc}(\rho(u,t))=\Dcal(u,t).
 \label{eq:catalyticrate}
\end{equation}
The same equality holds for correlated-catalytic LOCC distillation.
\end{corollary}
\begin{proof}
For each finite \(k\), the regularized quantity
\(D_{\PPT_k}^\infty\) is additive, strongly superadditive, asymptotically
continuous, and monotone under the correlated catalytic transformations above
\cite{Ao2026}. Those properties provide a rate bound independent of the
catalyst: additivity evaluates the independent input, strong superadditivity
separates the correlated output marginals, and the exactly returned catalyst
contribution cancels. Writing $E_k=D_{\PPT_k}^{\infty}$, an input catalyst $\tau_C$ and
output $\omega_{S'C}$ satisfy
\begin{align}
 n\mathcal D+E_k(\tau_C)
 &=E_k(\rho^{\otimes n}\otimes\tau_C)
 \ge E_k(\omega_{S'C})\notag\\
 &\ge E_k(\omega_{S'})+E_k(\omega_C)
 =E_k(\omega_{S'})+E_k(\tau_C).
 \label{eq:catalystcancellation}
\end{align}
The first inequality is PPT monotonicity and the second is strong
superadditivity. Exact recovery of the catalyst marginal cancels its
contribution even if its dimension grows; continuity is needed only for
the target system. If that system approaches $m$ Bell pairs, asymptotic
continuity gives $E_k(\omega_{S'})\ge m-o(n)$ for a finite proposed rate.

On \(n\) input copies its system contribution is
\(n\Dcal\), by Eq.~\eqref{eq:exactDn}. On \(m\) target Bell pairs it is \(m\).
The output approximation contributes only \(o(n)\) for any finite proposed
rate, by asymptotic continuity on the system output. Therefore
\(m/n\le\Dcal+o(1)\). No catalyst is required for the hashing protocol attaining
\(\Dcal\), so the bound is tight for both operation classes.
\end{proof}
This applies the catalytic monotone to the present NPT family. It does not
compute a catalytic formation cost or establish catalytic irreversibility.
Catalytic irreversibility of general higher-dimensional mixed states
\cite{Ao2026} therefore does not follow for Eq.~\eqref{eq:rho}.

\subsection{A measured key rate and a filtering comparison}
For \(c>0\), take a purification
\begin{equation}
 \ket{\Omega}=\sqrt{p_+}\ket{00}\ket{e_0}
             +\sqrt{p_-}\ket{11}\ket{e_1},\qquad
 \langle e_1|e_0\rangle=\eta=t/c.
 \label{eq:purification}
\end{equation}
Measuring both qubits in \(Z\) gives perfectly matching raw bits with entropy
\(\hbin(p_+)\). The conditional environment states are pure, and their average
has entropy \(S(\rho)\). The one-way privacy-amplification rate of this specified raw-key
protocol \cite{DevetakWinter} is therefore
\begin{equation}
 K_Z=I(Z_A;Z_B)-\chi(Z_A:E)=\hbin(p_+)-S(\rho)=\Dcal(u,t).
 \label{eq:key}
\end{equation}
The average environment state is
$\rho_E=p_+\ket{e_0}\bra{e_0}+p_-\ket{e_1}\bra{e_1}$.
Its determinant on its two-dimensional span is
$p_+p_-(1-\eta^2)=(1-u^2-t^2)/4$, so its eigenvalues are
$(1\pm s)/2$. Conditional on the $Z$ outcome the environment is pure,
and hence $\chi(Z_A:E)=S(\rho_E)=h_2((1+s)/2)$.
Bob's bit is Alice's bit without error, giving
$I(Z_A;Z_B)=h_2(p_+)$. This computes each term of Eq.~\eqref{eq:key}.

This is an achievable protocol rate, not a claim about the globally optimized
secret-key capacity.

An elementary alternative first applies Alice's diagonal filter
\begin{equation}
 F_A=\sqrt{\frac{p_-}{p_+}}\ket0\bra0+\ket1\bra1.
 \label{eq:filter}
\end{equation}
Since $p_-\le p_+$, the success Kraus operator satisfies
$F_A^\dagger F_A\le I$; the other outcome can be chosen as
$(I-F_A^\dagger F_A)^{1/2}$. The unnormalized success block is
\begin{equation}
 (F_A\otimes I)\rho(F_A\otimes I)
 =\begin{pmatrix}p_-&\eta p_-\\\eta p_-&p_-\end{pmatrix}_{00,11}.
 \label{eq:filterblock}
\end{equation}
Its trace is $2p_-$, and normalization gives diagonal entries $1/2$
and off-diagonal entries $\eta/2$.

For \(u<1\) it succeeds with probability \(2p_-=1-u\), producing
\(\rho(0,\eta)\). Its subsequent hashing rate per initial copy is
\begin{equation}
 R_{\mathrm{filter}}=(1-u)
 \left[1-\hbin\!\left(\frac{1+\eta}{2}\right)\right]\le\Dcal(u,\eta c).
 \label{eq:filterrate}
\end{equation}
The inequality follows from the optimality in Eq.~\eqref{eq:exactEd}, since
filtering followed by hashing is an LOCC protocol. For \(\eta=1\), the filter
produces a Bell pair on success; with dephasing it instead produces a balanced
mixed state. Filtering discards a fraction of the inputs, so its conditional
quality and its rate per original copy must be distinguished.

\begin{figure}[htbp]
\centering
\includegraphics[width=0.75\linewidth]{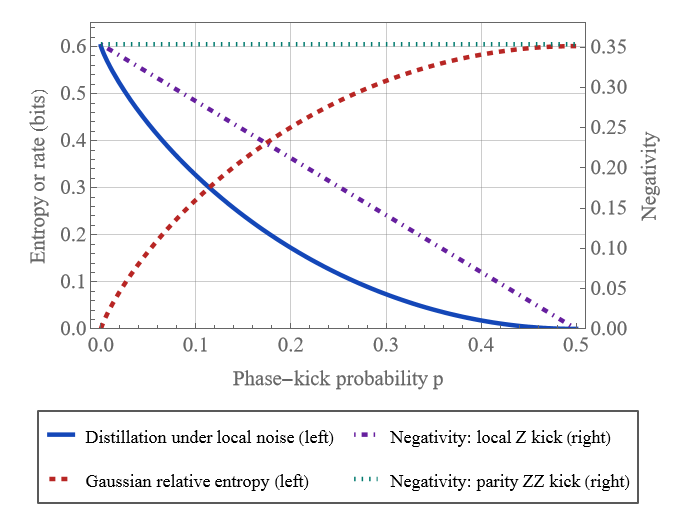}
\caption{Local phase noise and protected parity noise at
$u=c=1/\sqrt2$. The horizontal coordinate is the kick probability $p$.
The left axis gives the local-noise distillation rate and matched-Gaussian
relative entropy in bits, whose sum is the initial entanglement entropy.
The right axis gives negativity under $Z_B$ noise and, for comparison,
under $Z_AZ_B$ noise of the same probability. The latter is exactly constant.}
\label{fig:noise}
\end{figure}

\rev{Figure~\ref{fig:noise} separates two effects that are not fixed by the
same kick probability alone.  Local phase noise continuously removes the
coherence that supports the distillable resource and transfers the missing
part into the matched-Gaussian relative-entropy contribution, whereas the
correlated parity kick leaves the supported state unchanged.  The constant
protected-noise negativity is therefore a consequence of the parity symmetry,
not a claim that arbitrary phase noise is harmless.}

\subsection{Equal state fidelity does not fix the distillation rate}
There is an exact comparison within the already defined preparation family.
Fix the target $(u_0,c_0)$ on the pure circle, and compare local dephasing
with isotropic contraction before the ideal CNOT in Appendix~\ref{app:gate}.
Their parameters and target fidelities are
\begin{align}
 \rho_{\mathrm{ph}}&=\rho(u_0,\eta c_0),&
 F_{\mathrm{ph}}&=\tfrac12(1+u_0^2+\eta c_0^2),\notag\\
 \rho_{\mathrm{prep}}&=\rho(r u_0,r c_0),&
 F_{\mathrm{prep}}&=\tfrac12(1+r).
 \label{eq:matchedFidelityStates}
\end{align}
For a specified common fidelity $f\in[(1+u_0^2)/2,1]$, solve
\begin{equation}
 r=2f-1,\qquad\eta=\frac{2f-1-u_0^2}{c_0^2}.
 \label{eq:matchedFidelityParameters}
\end{equation}
Both states are physical, but their exact distillation rates are
\begin{align}
 D_{\mathrm{ph}}(f)&=h_2\!\left(\frac{1+u_0}{2}\right)
 -h_2\!\left(\frac{1+\sqrt{u_0^2+c_0^2\eta^2}}2\right),\notag\\
 D_{\mathrm{prep}}(f)&=h_2\!\left(\frac{1+r u_0}{2}\right)-h_2(f).
 \label{eq:matchedFidelityRates}
\end{align}
At the lower common-fidelity endpoint, $\eta=0$ and $D_{\mathrm{ph}}=0$.
The contraction state instead has $r=u_0^2$ and $t=r c_0>0$ for
$0<u_0<1$, so its distillation rate is strictly positive. Equality of
fidelity has not removed the different populations and coherences.
The noiseless endpoint gives the same original state and rate. No ordering
of these two rates for all other input families follows from this example.

\begin{figure}[htbp]
\centering
\includegraphics[width=0.75\linewidth]{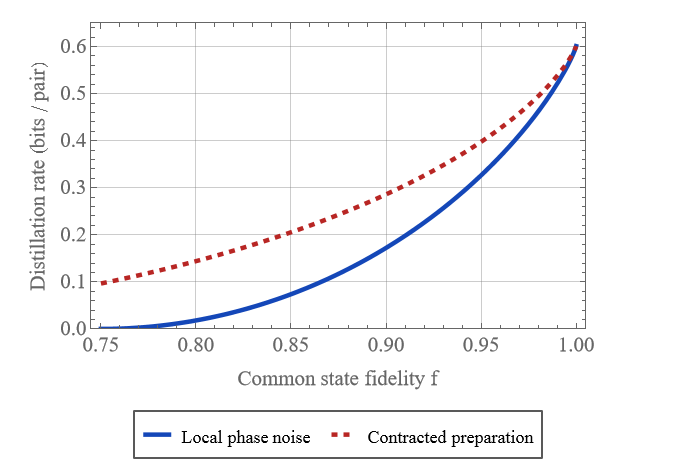}
\caption{Two calibrated noisy preparations with the same fidelity to the
original pure state at $u_0=c_0=1/\sqrt2$. The two rates are calculated from
Eq.~\eqref{eq:matchedFidelityRates}. At $f=3/4$, local dephasing is separable
but the contracted preparation remains distillable. Both coincide at $f=1$.
The horizontal coordinate is state fidelity, not an ensemble-averaged gate
fidelity or a no-error trajectory probability.}
\label{fig:matchedfidelity}
\end{figure}

\rev{Figure~\ref{fig:matchedfidelity} shows directly that fidelity to the same
target does not order the operational entanglement of two noisy preparations.
At the common lower endpoint, dephasing has already produced a separable
state, whereas the contracted preparation retains nonzero coherence and a
strictly positive distillation rate.  Their agreement at unit fidelity is only
the common noiseless endpoint and does not imply equivalence away from it.}

\section{Pauli probabilities, stabilizer work, and Gaussian resources}
\label{sec:resourceextensions}
\subsection{Bell participation probabilities rather than a CFT limit}
The doubled-space Bell participation identity \cite{HoshinoCFT} supplies an operational representation of the pure-state
Pauli moments. For an $n$-qubit pure state, write $d=2^n$ and
$\ket{B_P}=d^{-1/2}\operatorname{vec}(P)$. Measuring this orthonormal
maximally entangled basis on $\ket\psi\otimes\ket{\psi^*}$ gives
\begin{equation}
 q_P=\left|\langle B_P|\psi\otimes\psi^*\rangle\right|^2
     =\frac{|\langle P\rangle_\psi|^2}{d},\qquad
 M_\alpha^S(\psi)=H_\alpha(q)-\log_2d.
 \label{eq:Bellparticipation}
\end{equation}
Complex conjugation is in the specified computational basis; it is not a
general physical operation available on an unknown input. Our canonical
state is real, so two identical prepared copies implement this doubled input.
The Bell measurements here couple corresponding qubits of different copies;
they are not the two spatial measurement choices in a CHSH experiment.

For $\ket{\psi_\gamma}$, the only nonzero probabilities are
\begin{equation}
 (q_{II},q_{ZI},q_{IZ},q_{ZZ},q_{XX},q_{YY})
 =\frac14(1,u^2,u^2,1,c^2,c^2).
 \label{eq:Bellprobtable}
\end{equation}
They sum to one, and give, for $\alpha>0$, $\alpha\ne1$,
\begin{equation}
 M_\alpha^S(\psi_\gamma)
 =\frac1{1-\alpha}\log_2\left[\frac{1+u^{2\alpha}+c^{2\alpha}}2\right],
 \qquad \sum_Pq_P^2=\frac{1-u^2+u^4}{4}.
 \label{eq:Bellcollision}
\end{equation}
Thus the previously calculated $M_2$ can be recovered from a collision
probability of a completely specified measured distribution. This measured identity is valid at finite system size. Conformal boundary conditions,
universal $g$ factors, and logarithmic mutual-SRE laws require a spatially
extended critical system; no central charge or conformal critical point is
assigned to the two-qubit deformation.

\subsection{An exact stabilizer partition function}
The stabilizer statistical-mechanics generating function \cite{Salazar} packages the even statistics of all Pauli expectations into
an auxiliary partition function. In its convention,
\begin{equation}
\begin{aligned}
& Z_\beta(\psi)=e^{-\beta}\sum_{P\in\mathcal P_n}\cosh(\beta\langle P\rangle)
  = e^{-\beta} \sum_{P} \frac{e^{\beta \vert{}\langle P \rangle\vert{}} + e^{-\beta \vert{}\langle P \rangle\vert{}}}{2},\\
 & Z_\beta^c=Z_\beta-4^n e^{-\beta},\\
& W_\beta=\log_2\frac{2^n e^{-\beta}(\cosh\beta-1)}{Z_\beta^c(\psi)}.
 \label{eq:SPFdefinition}
 \end{aligned}
\end{equation}
The labels of Pauli strings are not retained by this generating function.
The parameter $\beta$ is auxiliary, not the inverse physical temperature of
$H_\gamma$. For $\beta>0$ the pure canonical state gives the closed expressions
\begin{align}
 Z_\beta(\psi_\gamma)
 &=e^{-\beta}[10+2\cosh\beta+2\cosh(\beta u)+2\cosh(\beta c)],
 \label{eq:SPFclosed}\\
 W_\beta(\psi_\gamma)
 &=\log_2\frac{2(\cosh\beta-1)}{
 (\cosh\beta-1)+(\cosh\beta u-1)+(\cosh\beta c-1)}.
 \label{eq:workclosed}
\end{align}
The multiplicities can be derived from all sixteen Pauli expectations,
without using a temperature limit. Work in the local-phase convention of
Eq.~\eqref{eq:uc} and put
\begin{equation}
 \ket\psi=a\ket{00}+b\ket{11},\quad
 a=\sqrt{p_+},\quad b=\sqrt{p_-},\quad
 a^2+b^2=1,\quad a^2-b^2=u,\quad2ab=c.
 \label{eq:PauliRealAmplitudes}
\end{equation}
For $0<|\gamma|<\infty$, $a>b>0$ and $0<u,c<1$. For negative
$\gamma$ this positive $b$ is the amplitude after the stated local phase
change, rather than the signed $\sinh\gamma/\sqrt{\cosh2\gamma}$ before it.
Using $X\ket0=\ket1$, $Y\ket0=i\ket1$, and
$Y\ket1=-i\ket0$ gives
\begin{align}
 II\ket\psi=ZZ\ket\psi&=a\ket{00}+b\ket{11},\notag\\
 ZI\ket\psi=IZ\ket\psi&=a\ket{00}-b\ket{11},\notag\\
 XX\ket\psi&=b\ket{00}+a\ket{11},\notag\\
 YY\ket\psi&=-b\ket{00}-a\ket{11},\notag\\
 XY\ket\psi=YX\ket\psi&=-ib\ket{00}+ia\ket{11}.
 \label{eq:PauliActionsDetailed}
\end{align}
The complete expectation table, with the row specifying Alice's operator
and the column Bob's, is therefore
\begin{equation}
 \big[\langle\sigma_A\otimes\sigma_B\rangle\big]_{\sigma_A,\sigma_B=I,X,Y,Z}
 =\begin{array}{c|rrrr}
   &I&X&Y&Z\\\hline
 I&1&0&0&u\\
 X&0&c&0&0\\
 Y&0&0&-c&0\\
 Z&u&0&0&1
 \end{array}.
 \label{eq:allPauliExpectations}
\end{equation}
Eight vanishing entries, $IX,IY,XI,YI,XZ,YZ,ZX,ZY$, reverse the
global parity and hence map the even support into its orthogonal odd
support. The remaining two, $XY$ and $YX$, preserve the even support;
their expectations vanish instead because $-iab+iab=0$ for real
amplitudes. Thus their zero values must not be explained as parity leakage.
The only unit-magnitude entries in the finite, nonzero pure family are
$II$ and $ZZ$.

To relate this count to stabilizer nullity \cite{MagicConcentration,Salazar},
let $\mathcal P_n=\{I,X,Y,Z\}^{\otimes n}$ denote the $4^n$
phase-free Hermitian Pauli labels. The signed stabilizer group is
\begin{equation}
 \mathcal S(\psi)=\{sP:P\in\mathcal P_n,\ s\in\{+1,-1\},
                          \ sP\ket\psi=\ket\psi\},\quad
 |\mathcal S|=2^r,\quad \nu(\psi)=n-r.
 \label{eq:nullityDefinitionDetailed}
\end{equation}
For a Hermitian Pauli $P^2=I$, its variance in a pure state obeys
\begin{equation}
 \|(P-\langle P\rangle)\ket\psi\|^2=1-\langle P\rangle^2.
 \label{eq:PauliVarianceStabilizer}
\end{equation}
Hence $|\langle P\rangle|=1$ if and only if $P$ has the state as a
$+1$ or $-1$ eigenvector. Exactly one sign then belongs to
$\mathcal S(\psi)$, so the unit-magnitude label count equals the
signed group order. In the present interior family,
\begin{equation}
 \mathcal S(\psi_\gamma)=\{II,ZZ\},\qquad
 r=\nu(\psi_\gamma)=1.
 \label{eq:familyNullityDetailed}
\end{equation}
This counts a deficit of independent Pauli stabilizers. It is not an
error-correction distance or a continuous measure of the amount of magic.
Here $ZZ$ expresses the conserved global parity; it is not two
independently conserved local parities of the deformed Hamiltonian.

The large-$\beta$ behavior can now be derived term by term. For any
$m_P=\langle P\rangle$ with $|m_P|\le1$,
\begin{equation}
 e^{-\beta}\cosh(\beta m_P)
 =\tfrac12 e^{-\beta(1-|m_P|)}
  +\tfrac12 e^{-\beta(1+|m_P|)}
 \longrightarrow
 \begin{cases}\tfrac12,&|m_P|=1,\\0,&|m_P|<1.\end{cases}
 \label{eq:PauliPartitionLimitTerm}
\end{equation}
For an arbitrary pure $n$-qubit state the finite Pauli sum therefore gives
\begin{equation}
 \lim_{\beta\to\infty}Z_\beta(\psi)
 =\lim_{\beta\to\infty}Z_\beta^c(\psi)
 =\frac{|\mathcal S(\psi)|}{2}=2^{n-\nu(\psi)-1}.
 \label{eq:partitionCountGeneral}
\end{equation}
It is the stabilizer work, rather than the partition function itself,
whose limit equals nullity:
\begin{equation}
 \lim_{\beta\to\infty}W_\beta(\psi)
 =\log_2\frac{2^{n-1}}{|\mathcal S(\psi)|/2}
 =n-\log_2|\mathcal S(\psi)|=\nu(\psi).
 \label{eq:workNullityGeneral}
\end{equation}
For our two-qubit interior family both $Z_\infty$ and $\nu$ happen
to equal one. This numerical equality is not their general relation.

In this family the full exponential expansion is particularly explicit:
\begin{align}
 Z_\beta(\psi_\gamma)
 &=1+e^{-2\beta}+10e^{-\beta}
   +e^{-\beta(1-u)}+e^{-\beta(1+u)}\notag\\
 &\hspace{25mm}+e^{-\beta(1-c)}+e^{-\beta(1+c)}.
 \label{eq:partitionExponentialDetailed}
\end{align}
Every term after the first vanishes at fixed $0<u,c<1$. Defining the
positive auxiliary gap $\delta_\gamma=\min\{1-u,1-c\}$ gives
$Z_\beta=1+O_\gamma(e^{-\beta\delta_\gamma})$.
Similarly write
\begin{align}
 q_\beta&=\frac{\cosh(\beta u)-1+\cosh(\beta c)-1}
                    {\cosh\beta-1}\ge0,\notag\\
 W_\beta&=1-\log_2(1+q_\beta),\qquad
 1-W_\beta=\frac{q_\beta}{\ln2}+O(q_\beta^2),\notag\\
 q_\beta&=O_\gamma(e^{-\beta\delta_\gamma}).
 \label{eq:workNullityConvergence}
\end{align}
This supplies the convergence mechanism and its rate, in addition to the
limiting integer. The gap is a gap in the auxiliary Pauli-expectation
statistics, not an energy gap of $H_\gamma$.

At the exact product endpoint, the signed stabilizers are
$\{II,ZI,IZ,ZZ\}$; at the positive-phase Bell endpoint they are
$\{II,XX,-YY,ZZ\}$. Both have order four and nullity zero, while
$Z_\infty=2$. Substitution into Eq.~\eqref{eq:workclosed} gives
$W_\beta=0$ at either endpoint for every $\beta>0$.
The nonuniform limits are thus explicit:
\begin{align}
 \lim_{|\gamma|\to\infty}\lim_{\beta\to\infty}W_\beta=1,
 &\qquad
 \lim_{\beta\to\infty}\lim_{|\gamma|\to\infty}W_\beta=0,\notag\\
 \lim_{|\gamma|\to0^+}\lim_{\beta\to\infty}W_\beta=1,
 &\qquad
 \lim_{\beta\to\infty}\lim_{|\gamma|\to0^+}W_\beta=0.
 \label{eq:workEndpointLimitsDetailed}
\end{align}
Indeed $1-u\sim2\gamma^2$ near the product endpoint, while
$1-c\sim2e^{-4|\gamma|}$ near the Bell endpoint. Reaching the interior
nullity plateau requires $\beta\delta_\gamma\gg1$, a condition that
cannot be uniform in either endpoint. No physical cooling or many-body
ground-state degeneracy is inferred from this auxiliary limit.

The small-$\beta$ coefficient follows independently from the same
closed formula. Its numerator is $\beta^2+\beta^4/12+O(\beta^6)$,
and its denominator is
$\beta^2+\beta^4(1+u^4+c^4)/24+O(\beta^6)$, because $u^2+c^2=1$.
Their ratio is $1+\beta^2u^2c^2/12+O(\beta^4)$; taking $\log_2$
divides the leading correction by $\ln2$. Together with
Eq.~\eqref{eq:workNullityGeneral}, this gives
\begin{equation}
 W_\beta=\frac{\beta^2u^2c^2}{12\ln2}+O(\beta^4),\qquad
 \lim_{\beta\to\infty}W_\beta=1
 \quad(0<|\gamma|<\infty).
 \label{eq:worklimits}
\end{equation}

\begin{proposition}[A common resource-profile extremum]
For every fixed $\beta>0$, $W_\beta(\psi_\gamma)$ is maximized at
$u^2=c^2=1/2$, the same deformation maximizing
$\Mcal_\gamma=M_2^S(\psi_\gamma)$.
\end{proposition}
\begin{proof}
With $x=u^2$, the varying part of the denominator in
Eq.~\eqref{eq:workclosed} is $f_\beta(x)+f_\beta(1-x)$, where
$f_\beta(x)=\cosh(\beta\sqrt{x})$. Its power series has nonnegative
coefficients, and its second derivative is strictly positive on $[0,1]$ for
$\beta>0$. Convexity gives the unique minimum at $x=1/2$.
\end{proof}
This is a fixed-Pauli-frame work profile. It is not an optimization over all
local bases and does not replace the definition of nonlocal magic.

\subsection{Pure Gaussianity and a noisy-state Wick defect}
Interpret the computational basis as the ordered occupation basis of two
fermionic modes, with $\ket{11}=f_1^\dagger f_2^\dagger\ket{\mathrm{vac}}$.
The pure state has the BCS form
\begin{equation}
 \ket{\psi_\gamma}=\sqrt{p_+}
 \exp\!\left(\sqrt{p_-/p_+}\,f_1^\dagger f_2^\dagger\right)
 \ket{\mathrm{vac}}.
 \label{eq:BCS}
\end{equation}
It is Gaussian. This occupation-space interpretation does not impose a new
local fermionic parity superselection rule on the qubit LOCC protocols above.
Convolution-based fermionic non-Gaussianity \cite{HoshinoFNG} is defined for pure
even-parity states by
\begin{equation}
 M_2^{\mathrm{fNG}}(\psi)=-\ln\nu_2,\quad
 \nu_2=\Tr\mathcal C(\rho)^2,\quad
 \mathcal C(\rho)=\Tr_2[U_{12}\rho^{\otimes2}U_{12}^\dagger],\quad
 U_{12}=\exp\!\left[\frac\pi8\sum_j\xi_j^{(1)}\xi_j^{(2)}\right],
 \label{eq:FNGdefinition}
\end{equation}
where Majorana operators on all copies mutually anticommute. The pure
Gaussian characterization of that convolution gives
\begin{equation}
 M_2^{\mathrm{fNG}}(\psi_\gamma)=0,\qquad
 M_2^{\NL}(\psi_\gamma)>0\quad(0<|\gamma|<\infty).
 \label{eq:resourceinequivalence}
\end{equation}
The first quantity uses natural logarithms and Gaussian free operations;
the second uses bits and an optimization of qubit stabilizer entropy over
local unitaries. The resource theories are different. Universal
Majorana bounds and SYK large-system transitions \cite{HoshinoFNG} do not give a phase
transition of this two-mode pure Gaussian family.

This distinction becomes sharper under the local noise already present in
our model. For $\rho(u,t)$, Wick factorization would require
\begin{equation}
 \langle n_1n_2\rangle=
 \langle n_1\rangle\langle n_2\rangle
 -|\langle f_1^\dagger f_2\rangle|^2+|\langle f_1f_2\rangle|^2.
\end{equation}
Instead its exact defect is
\begin{equation}
 \Delta_{\mathrm{Wick}}=
 \langle n_1n_2\rangle-\langle n_1\rangle\langle n_2\rangle
 +|\langle f_1^\dagger f_2\rangle|^2-|\langle f_1f_2\rangle|^2
 =\frac{1-u^2-t^2}{4}=\frac{c^2(1-\eta^2)}4.
 \label{eq:Wickdefect}
\end{equation}
Here $\langle n_1\rangle=\langle n_2\rangle=p_-$,
$\langle n_1n_2\rangle=p_-$, the normal cross correlator is zero, and the
pairing magnitude is $t/2$. Therefore every genuinely mixed interior state
with $u^2+t^2<1$ is non-Gaussian, despite its Gaussian pure boundary.

One can quantify this with a distinct mixed-state relative-entropy defect.
Let $s=\sqrt{u^2+t^2}$, $q=(1+s)/2$, and $2\theta=\operatorname{atan2}(t,u)$.
The Gaussian state with the same covariance is obtained by applying the
BCS rotation to the product thermal state with eigenvalues
$q^2,q(1-q),q(1-q),(1-q)^2$. Its entropy is $2h_2(q)$, and the original
state's entropy is $h_2(q)$. Direct evaluation yields
\begin{equation}
 D(\rho(u,t)\|\rho_G(u,t))=S(\rho_G)-S(\rho)=h_2(q).
 \label{eq:Gaussianrelative}
\end{equation}
The matching Gaussian density matrix can also be constructed explicitly.
Let $U_\theta$ rotate the even basis by
\begin{equation}
 U_\theta\ket{00}=\cos\theta\ket{00}+\sin\theta\ket{11},\quad
 U_\theta\ket{11}=-\sin\theta\ket{00}+\cos\theta\ket{11},
 \label{eq:BCSrotation}
\end{equation}
and act as identity on the odd sector. This is a quadratic fermionic BCS
unitary. Then
\begin{align}
 U_\theta^\dagger\rho U_\theta&=\diag(q,0,0,1-q),\notag\\
 \rho_G&=U_\theta\diag(q^2,q(1-q),q(1-q),(1-q)^2)U_\theta^\dagger.
 \label{eq:Gaussianexplicit}
\end{align}
The two states have the same two-point covariance: the unrotated occupation
of either mode is $1-q$ in both, and the BCS unitary rotates their covariance
matrices identically. Their quartic correlations differ. Relative entropy
is unitary invariant, so evaluating the two nonzero input eigenvalues gives
\begin{align}
 D(\rho\|\rho_G)
 &=q\log_2\frac q{q^2}+(1-q)\log_2\frac{1-q}{(1-q)^2}
 =h_2(q).
 \label{eq:Gaussiantraceproof}
\end{align}
At $s=0$ the rotation angle can be chosen arbitrarily; $\rho_G=I/4$.
At $s=1$ the states coincide and the zero eigenvalue terms are interpreted
by continuity. This explicit construction avoids treating the Gaussian
relative-entropy expression as a renamed stabilizer entropy.

This is in bits, and is not Eq.~\eqref{eq:FNGdefinition} applied to a mixed
state. The matched covariance does not erase the quartic parity correlation.
Along fixed-population dephasing, this Gaussian defect obeys the exact
entropy budget
\begin{equation}
 \mathcal D(u,\eta c)+D(\rho(u,\eta c)\|\rho_G(u,\eta c))=h_2(p_+).
 \label{eq:entropybudget}
\end{equation}
The loss of distillable entanglement equals the increase of this particular
non-Gaussian relative entropy. This follows by adding two explicitly evaluated
entropies, not by postulating a resource-conversion law between Gaussian and
stabilizer theories. Figure~\ref{fig:noise} compares the two complementary
quantities with the protected and unprotected negativities.

\section{Sharing the deformed state through a swapping network}
\label{sec:swapping}
Target-matched entanglement swapping \cite{HuhtanenSwap}
can be specialized to the two nonzero Schmidt probabilities $p_+,p_-$.
Its intermediary makes this a three-party protocol whose conditioned output
is bipartite; the network is not a two-party Bell experiment.
Consider two identical pure links on $A E_1$ and $E_2 B$, and the
intermediary's target-matched projection
$\bra{\psi_\gamma}_{E_1E_2}$. Equivalently it applies the inverse state
preparation and heralds the specified computational outcome. Contraction gives
\begin{equation}
 (\id\otimes\bra{\psi_\gamma}\otimes\id)
 (\ket{\psi_\gamma}_{AE_1}\otimes\ket{\psi_\gamma}_{E_2B})
 =p_+^{3/2}\ket{00}_{AB}+p_-^{3/2}\ket{11}_{AB}.
 \label{eq:swapcontraction}
\end{equation}
The success probability, conditional squared fidelity to the target, and
conditional state parameters therefore are
\begin{align}
 P_0&=p_+^3+p_-^3=\frac{1+3u^2}{4},\label{eq:swapP}\\
 F_{\mathrm{sq}}&=\frac{(p_+^2+p_-^2)^2}{p_+^3+p_-^3}
 =\frac{(1+u^2)^2}{1+3u^2}\ge\frac89,\label{eq:swapF}\\
 u_{\mathrm{out}}&=\frac{u(3+u^2)}{1+3u^2},\qquad
 c_{\mathrm{out}}=\frac{c^3}{1+3u^2}.
 \label{eq:swapuc}
\end{align}
Equation~\eqref{eq:swapF} uses squared fidelity throughout. Its minimum occurs at $u^2=1/3$.
With $x=u^2$, differentiation of the squared fidelity gives
\begin{equation}
 \frac{d}{dx}\frac{(1+x)^2}{1+3x}
 =\frac{(1+x)(3x-1)}{(1+3x)^2}.
 \label{eq:swapextremum}
\end{equation}
The minimum on $[0,1]$ is at $x=1/3$, where the value is $8/9$.
The conditional amplitudes are $p_\pm^{3/2}/\sqrt{Z_3}$; their squared
difference gives $u_{\mathrm{out}}$, and twice their product gives
$2(p_+p_-)^{3/2}/Z_3=c^3/(1+3u^2)$.
Also $u_{\mathrm{out}}-u=2u(1-u^2)/(1+3u^2)>0$ in the interior.

For $0<u<1$, $u_{\mathrm{out}}>u$, so successful sharing with identical
target-matched resources reduces the entanglement entropy even though its
fidelity remains high. Its conditional nonlocal magic is obtained by
substituting $u_{\mathrm{out}}$ into Eq.~\eqref{eq:magic}. Fidelity alone
therefore does not establish preservation of either resource.

For $k\ge1$ such intermediaries, write $Z_j=p_+^j+p_-^j$.
Successive target-matched projections give
\begin{equation}
 p_\pm^{(k)}=\frac{p_\pm^{2k+1}}{Z_{2k+1}},\quad
 P_{\mathrm{all}}^{(k)}=Z_{2k+1},\quad
 F_{\mathrm{sq}}^{(k)}=\frac{Z_{k+1}^2}{Z_{2k+1}},\quad
 u_k=\tanh[(2k+1)\atanh u].
 \label{eq:swapnetwork}
\end{equation}
For the induction, after $j-1$ successful intermediary projections the
amplitudes are $p_i^{j-1/2}/\sqrt{Z_{2j-1}}$. Appending one original
link multiplies them by $p_i^{1/2}$ and projecting on the matched vector
multiplies them by another $p_i^{1/2}$. The unnormalized new amplitudes
are $p_i^{j+1/2}/\sqrt{Z_{2j-1}}$, with conditional success probability
$Z_{2j+1}/Z_{2j-1}$. The overlap with the original target is
$Z_{j+1}/\sqrt{Z_{2j+1}}$, proving the fidelity and the full recursion.

Here the success probability is for all the specified outcomes: the individual
conditional factors $Z_{2j+1}/Z_{2j-1}$ telescope, with $Z_1=1$.
For fixed nonzero $u$, conditioned states tend to a product as $k$ increases,
while the success probability decays. At $u=0$ the conditional target is a
Bell pair, but selecting one outcome at every node still costs $4^{-k}$.

The flattened-spectrum alternative has a different resource accounting.
For one node, replace the second link by a Bell pair and use a complete
Bell measurement at the node. All four outcomes occur with probability $1/4$;
after the communicated Pauli correction on Bob's qubit, the output is exactly
$\ket{\psi_\gamma}$. This implements deterministic sharing of an arbitrary
Schmidt-weighted target by consuming a flat-spectrum link. It does not make
two identical nonmaximal links into a deterministic exact target for free.
The herald or correction record is required; averaging over uncommunicated
outcomes does not create signalling between endpoints.

\rev{The flat-spectrum construction also extends to an arbitrary number
\(k\) of intermediate nodes \cite{HuhtanenSwap}. Specializing its common
Schmidt basis to the present rank-two family, the intermediate flattened
resources are Bell links. Complete Bell measurements and communicated Pauli
frame corrections then transport the nonuniform endpoint amplitudes without
the power-law sharpening in Eq.~\eqref{eq:swapnetwork}. After the final
correction one has}
\begin{equation}
 \rev{\rho_{AB}^{(k,\mathrm{flat})}
 =\ket{\psi_\gamma}\!\bra{\psi_\gamma},\qquad
 F_{\mathrm{sq}}^{(k,\mathrm{flat})}=1,\qquad
 P_{\mathrm{total}}^{(k,\mathrm{flat})}=1.}
 \label{eq:flatSwapNetwork}
\end{equation}
\rev{Thus exact deterministic sharing at arbitrary depth is available, but
its resource accounting uses flat-spectrum Bell links, complete measurement
records, and classical feedforward. It does not convert a chain of identical
nonmaximally entangled links into the target state for free.}

\section{A finite-sample certificate within the calibrated model}
\label{sec:certificate}
Suppose the real maximally correlated support in Eq.~\eqref{eq:rho} is known
from independent preparation calibration, with iid copies and trusted Pauli
measurements. The two parameters can then be measured as
\begin{equation}
 u=\langle Z\otimes I\rangle= \operatorname{Tr}[\rho_{AB}(Z \otimes I_B)] = \operatorname{Tr}_A[\operatorname{Tr}_B(\rho_{AB}) Z] = \operatorname{Tr}(\rho_A Z) = \operatorname{Tr}(Z\rho_A),\qquad t=\langle X\otimes X\rangle,
 \label{eq:twosettings}
\end{equation}
up to the local sign convention already fixed. Use \(N\) copies for each
observable and denote the sample averages by \(\hat u,\hat t\). For a failure
probability \(\epsilon\in(0,1)\), define
\begin{equation}
 r_N=\sqrt{\frac{2\ln(4/\epsilon)}N},\qquad
 u_L=\max\{0,|\hat u|-r_N\},\quad
 t_L=\max\{0,|\hat t|-r_N\}.
 \label{eq:confidence}
\end{equation}
Hoeffding's inequality for \(\pm1\)-valued observations and a union bound give
simultaneous parameter coverage with probability at least \(1-\epsilon\).
The rectangle defined by these intervals must intersect the physical disk
\(u^2+t^2\le1\); otherwise the model and data are inconsistent at this
confidence level, and the following certificate is not issued.

\begin{proposition}[Model-constrained resource bound]
Under these assumptions, with probability at least \(1-\epsilon\),
\begin{equation}
 E_d(\rho)\ge\Dcal(u_L,t_L),\qquad \mathcal N(\rho)\ge t_L/2.
 \label{eq:certbound}
\end{equation}
\end{proposition}
\begin{proof}
On the coverage event, \(u_L\le u,t_L\le t\), so the lower pair is physical.
In the interior, differentiating Eq.~\eqref{eq:introD} gives
\begin{equation}
 \partial_t\Dcal=\frac{t}{s\ln2}\atanh s\ge0,\qquad
 \partial_u\Dcal=\frac{(u/s)\atanh s-\atanh u}{\ln2}\ge0.
 \label{eq:derivatives}
\end{equation}
The second inequality follows because \(\atanh z/z\) is increasing for
\(0<z<1\) and \(s\ge u\). Both inequalities extend to endpoints by continuity.
Reducing either coordinate while staying in the disk cannot increase
\(\Dcal\). The distillation identity and Eq.~\eqref{eq:neg} complete the proof.
\end{proof}
For each $\pm1$ sample average, Hoeffding's bound is
$P(|\hat u-u|>r)\le2e^{-Nr^2/2}$, and likewise for $t$. A union bound
is at most $4e^{-Nr^2/2}$; setting it to $\varepsilon$ gives
Eq.~\eqref{eq:confidence}. The monotonicity used in the rate bound is
also explicit in the convergent series
\begin{equation}
 \frac{\operatorname{artanh}z}{z}
 =\sum_{j=0}^{\infty}\frac{z^{2j}}{2j+1},\qquad0\le z<1.
 \label{eq:monotonicityseries}
\end{equation}
Each nonconstant term is increasing. Thus
\begin{equation}
 (u/s)\operatorname{artanh}s-\operatorname{artanh}u
 =u[\operatorname{artanh}s/s-\operatorname{artanh}u/u]\ge0.
 \label{eq:monotonicityproof}
\end{equation}
The path from $(u_L,t_L)$ to $(u,t)$ increasing one coordinate at a time
stays in the physical disk because both endpoint coordinates are bounded
by the physical upper pair. No extrapolation outside the disk is required.

Only two observables are required after assuming this state family. They do not
certify its support, reality, absence of leakage, or iid structure. The bound is
neither device independent nor robust to arbitrary SPAM errors, and it is a
confidence bound on an asymptotic rate, not a finite-block guarantee for the
number or fidelity of distilled Bell pairs. Universal unknown-state distillation \cite{Takagi2026} supplies a separate existence theorem for unknown
iid states; that result should not be replaced by a naive plug-in estimate of
Eq.~\eqref{eq:Dgammaeta}.

\section{Asymptotic regularization and collective nonlocal magic}
\label{sec:regularization}
\subsection{Optimized coherent information and PPT comparison distances}
Many-copy regularization evaluates a specified functional on tensor powers
and divides by the number of copies. For optimized coherent information
\cite{Takagi2026} and PPT comparison distances \cite{Ao2026}, respectively,
\begin{align}
 E_d^{\le}(\rho)&:=\sup_{\Lambda\in\LOCC}
       [S(B')_{\Lambda(\rho)}-S(A'B')_{\Lambda(\rho)}],\notag\\
 E_d(\rho)&=\lim_{m\to\infty}\frac1m E_d^{\le}(\rho^{\otimes m}),
 \label{eq:Takagi9}\\
 D_{\PPT_k}^\infty(\rho)&:=\lim_{m\to\infty}\frac1m
 D_{\PPT_k(A^m:B^m)}(\rho^{\otimes m}).
 \label{eq:Ao9}
\end{align}
The optimization in the first line includes LOCC preprocessing and its
output systems $A',B'$; it is not a finite-error one-shot distillation yield.
The second construction uses the infimum over comparison operators in
Eq.~\eqref{eq:Rains}. The shared per-copy limiting structure does not identify
the two parent functionals or their operational meanings.

For the maximally correlated family the equality can be proved at every
finite block size. Doing nothing before hashing gives
$E_d^{\le}(\rho^{\otimes m})\ge m\mathcal D$. Conversely, for any
LOCC preprocessing $\Lambda$, hashing of its iid output achieves at least
$I_c(\Lambda(\rho^{\otimes m}))$ per block. Preprocessing followed by
hashing is an allowed LOCC protocol on the original blocks, so
\begin{equation}
 I_c(\Lambda(\rho^{\otimes m}))\le E_d(\rho^{\otimes m})=m\mathcal D.
 \label{eq:coherentupper}
\end{equation}
Taking the supremum and using Eq.~\eqref{eq:exactDn} yields
\begin{equation}
 E_d^{\le}(\rho^{\otimes m})=D_{\PPT_k}(\rho^{\otimes m})=m\mathcal D,
 \quad E_d(\rho)=D_{\PPT_k}^{\infty}(\rho)=\mathcal D.
 \label{eq:equalregularands}
\end{equation}
The equality follows from the matching hashing and comparison bounds, not
from regularization alone. Additivity of a regularized distance is also
distinct from additivity of its unregularized parent on general states.

\subsection{Pairwise and collective local-unitary domains}
Fix $0<|\gamma|<\infty$ and let
$\ket{\Psi_\gamma^{(m)}}=\ket{\psi_\gamma}^{\otimes m}$ with the cut
$A^m:B^m$. Its Schmidt expansion is
\begin{equation}
 \ket{\Psi_\gamma^{(m)}}=\sum_{x\in\{0,1\}^m}
  \sqrt{p_+^{m-|x|}p_-^{|x|}}\ket{x}_{A^m}\ket{x}_{B^m}.
 \label{eq:manySchmidt}
\end{equation}
The squared Schmidt value $p_+^{m-j}p_-^j$ occurs $\binom mj$ times.
This binomial spectrum supplies the many-copy input; the Hamiltonian,
positive inner product, and original one-pair state are unchanged.

For $2m$ qubits the fixed-frame definition is
\begin{equation}
 \zeta_4(\Psi)=\frac1{4^m}\sum_{P\in\mathcal P_{2m}}
                    \langle\Psi|P|\Psi\rangle^4,\qquad
 M_2^S(\Psi)=-\log_2\zeta_4(\Psi).
 \label{eq:manySREdefinition}
\end{equation}
Every Pauli string factors into $m$ two-qubit strings. Tensor-product
expectations and the finite sums therefore factorize:
\begin{align}
 \zeta_4(\psi_\gamma^{\otimes m})
 &=\prod_{j=1}^m\left[\frac14\sum_{P_j\in\mathcal P_2}
                         \langle P_j\rangle_{\psi_\gamma}^4\right]
 =(1-u_\gamma^2+u_\gamma^4)^m,\notag\\
 M_2^S(\psi_\gamma^{\otimes m})&=m\mathcal M_\gamma.
 \label{eq:fixedmagic}
\end{align}
Now distinguish the two local-unitary optimizations explicitly:
\begin{align}
 \mathcal M_m^{\mathrm{pair}}&:=
 \min_{\{U_{A,j},U_{B,j}\in U(2)\}}
 M_2^S\left(\bigotimes_{j=1}^m
             (U_{A,j}\otimes U_{B,j})\psi_\gamma\right),
 \label{eq:pairdomain}\\
 \mathcal M_m^{\mathrm{col}}&:=
 \min_{U_A,U_B\in U(2^m)}
 M_2^S[(U_A\otimes U_B)\Psi_\gamma^{(m)}]
 =M_2^{\NL}(\Psi_\gamma^{(m)}).
 \label{eq:collectivedomain}
\end{align}
In the first domain the output is a product across copies. Additivity lets
the minimum be taken independently in each factor; the rank-two
Schmidt-frame optimum \cite{Sierant} then gives
\begin{equation}
 \mathcal M_m^{\mathrm{pair}}=m\mathcal M_\gamma,\qquad
 0\le\mathcal M_m^{\mathrm{col}}\le\mathcal M_m^{\mathrm{pair}}.
 \label{eq:domainscomparison}
\end{equation}
The inequality follows from inclusion of the pairwise domain in the
collective domain. Collective operations remain local across $A^m:B^m$,
but can couple different copies within either laboratory.

\subsection{Positive pairwise density and vanishing collective density}
The imperfect-dimer spectral asymptotics \cite{Sierant} applied to
Eq.~\eqref{eq:manySchmidt} give, for fixed nonzero finite deformation,
\begin{equation}
 \mathcal M_m^{\mathrm{col}}=\frac32\log_2m+O_\gamma(1).
 \label{eq:manymagic}
\end{equation}
The exponent can be traced to the binomial Schmidt spectrum rather than
postulated as a fit. Write $p=p_+$, $q=p_-$, so $1/2<p<1$ and $q$ is
fixed. Its level masses are
\begin{equation}
 b_j=\binom mjp^{m-j}q^j,\qquad
 \frac{p^{m-j-1}q^{j+1}}{p^{m-j}q^j}=\frac qp,
 \quad j=0,\ldots,m.
 \label{eq:binomialmasses}
\end{equation}
For $j=mq+z\sqrt{mpq}$ in the central window, Stirling's formula gives
\begin{equation}
 b_j=\frac{e^{-z^2/2}}{\sqrt{2\pi mpq}}[1+o(1)].
 \label{eq:binomiallocal}
\end{equation}
The mass is spread across order $\sqrt m$ levels of order $m^{-1/2}$.
The fourth-moment sum has the corresponding Riemann-sum asymptotics
\begin{align}
 \sum_j b_j^4
 &\sim\frac{\sqrt{mpq}}{(2\pi mpq)^2}
       \int_{-\infty}^{\infty}e^{-2z^2}\,dz\notag\\
 &=\frac{1}{4\sqrt2\,\pi^{3/2}(mpq)^{3/2}}.
 \label{eq:binomialfourth}
\end{align}
The Gaussian approximation applies on enlarging central windows; binomial
tail estimates control the remaining levels when passing to this moment.

To connect level masses to a basis-independent magic bound, sort all
Schmidt probabilities as $\lambda_0\ge\lambda_1\ge\cdots$ and form
dyadic rank shells $D_0=\{0\}$ and
$D_k=\{2^{k-1},\ldots,2^k-1\}$. Define the shell envelopes
\begin{equation}
 \widehat w_0=\lambda_0,\quad
 \widehat w_k=2^{k-1}\lambda_{2^{k-1}},\quad
 Q_m=\sum_{k=0}^{m}\widehat w_k^4.
 \label{eq:dyadicenvelopes}
\end{equation}
The envelope bounds each shell's mass from above. The spectral shell
comparison \cite{Sierant} controls the binomial levels and their tails by
constants depending on $p$, giving
$Q_m=\Theta_p(\sum_jb_j^4)=\Theta_p(m^{-3/2})$.
In the central region this comparison has a concrete rank interpretation:
the cumulative multiplicity $N_j=\sum_{l\le j}\binom ml$ is comparable
to $\binom mj$, and successive $N_j$ have a ratio bounded away from one
and infinity for fixed $q<1/2$. A dyadic rank shell therefore samples only
a bounded number of central binomial levels, and a level occupies only
a bounded number of such shells. The fixed probability ratio $q/p$
keeps the corresponding envelope and level masses comparable. The
uniform spectral purity estimate then gives
\begin{equation}
 M_{\mathrm{CB}}(m)=-\log_2Q_m+O(1),\qquad
 0\le M_{\mathrm{CB}}(m)-\mathcal M_m^{\mathrm{col}}\le4.
 \label{eq:shellmagiccomparison}
\end{equation}
Taking $-\log_2$ of the $m^{-3/2}$ shell moment produces
$(3/2)\log_2m$, and the uniformly bounded optimization gap changes only
the constant term. This proves how the many-copy scaling follows from
the binomial input and the spectral shell estimates. The constants in
that comparison are not uniform as $p$ approaches $1/2$ or one.

The finite-copy spectral calculation below checks the same binomial input;
it is not a derivation of the asymptotic theorem from a short numerical fit.
For sufficiently large $m$, the remainder has magnitude at most a constant
$K_\gamma$ independent of $m$. Consequently
\begin{equation}
 0\le\frac{\mathcal M_m^{\mathrm{col}}}{m}
 \le\frac{\frac32\log_2m+K_\gamma}{m}\longrightarrow0.
 \label{eq:magicdensitysqueeze}
\end{equation}
The two regularized densities are thus
\begin{align}
 \mathcal M_{\mathrm{pair}}^\infty(\gamma)
 &:=\lim_{m\to\infty}\frac{\mathcal M_m^{\mathrm{pair}}}{m}
 =-\log_2(1-u_\gamma^2+u_\gamma^4)>0,\notag\\
 \mathcal M_{\mathrm{col}}^\infty(\gamma)
 &:=\lim_{m\to\infty}\frac{\mathcal M_m^{\mathrm{col}}}{m}=0,
 \qquad\frac{\mathcal M_m^{\mathrm{col}}}{\mathcal M_m^{\mathrm{pair}}}
 \longrightarrow0.
 \label{eq:contrast}
\end{align}
The common regularization viewpoint \cite{Takagi2026,Ao2026} makes the
operation-domain dependence visible: the fixed or pairwise optimized
stabilizer resource is extensive, whereas the collectively optimized
nonlocal part is logarithmically subextensive. Subadditivity alone would
not prove zero density; the sublinear estimate in Eq.~\eqref{eq:manymagic}
is the essential extra input.

Entanglement behaves differently. The reduced state is
$\rho_A^{\otimes m}$, whose entropy is $m h_2(p_+)$. A collective local
unitary only conjugates that reduced matrix, leaving its spectrum unchanged.
Thus
\begin{equation}
 \frac1m E_d(\psi_\gamma^{\otimes m})
 =\frac1mD_{\PPT_k}(\psi_\gamma^{\otimes m})
 =h_2(p_+)>0.
 \label{eq:entdensitycontrast}
\end{equation}
The zero collective magic density coexists with positive entanglement
density without destroying the entanglement. It is not a zero operational
magic-distillation rate: arbitrary local unitaries are not generally
stabilizer-free operations, and Eq.~\eqref{eq:collectivedomain} specifies
no magic conversion protocol. The limit is a copy-number limit rather than
a spatial phase transition. At the exact product and Bell endpoints every
copy is stabilizer and both magic quantities vanish; Eq.~\eqref{eq:manymagic}
is not uniform in those endpoints. A joint $\gamma$--$m$ crossover cannot
be inferred by interchanging these limits.

\subsection{\rev{Stabilizer-fidelity nonlocal magic and entanglement capacity}}
\rev{A complementary nonlocal-magic monotone is obtained by minimizing the
negative logarithm of stabilizer fidelity over arbitrary local basis changes
\cite{NonlocalMagicSpreading}.  For an ordered Schmidt spectrum
$\{\lambda_j\}$, define the prefix fidelities
$F_r=r^{-1}(\sum_{j=1}^{r}\sqrt{\lambda_j})^2$, where the admissible
stabilizer Schmidt ranks are $r=2^k$.  In the present rank-two problem this
leaves only the best product and maximally entangled stabilizer
approximations:}
\begin{align}
 \rev{F_1}&\rev{=p_+=\frac{1+u}{2},}\notag\\
 \rev{F_2}&\rev{=\frac{(\sqrt{p_+}+\sqrt{p_-})^2}{2}
              =\frac{1+c}{2},}\notag\\
 \rev{\mathfrak D_{\rm NL}(\psi_\gamma)}
 &\rev{=-\log_2\!\left[\frac{1+\max\{u,c\}}{2}\right].}
 \label{eq:onecopyFidelityNL}
\end{align}
\rev{Thus this measure vanishes at both stabilizer endpoints and is largest
where $u=c=1/\sqrt2$.  It is not the order-two stabilizer R\'enyi entropy
$\mathcal M_\gamma$ used above.}

\rev{For $m$ copies the entanglement energy is a sum of independent binary
variables.  Its base-two variance, or capacity of entanglement, is therefore}
\begin{equation}
 \rev{C_m=m C_1,\qquad
 C_1=p_+p_-\!\left[\log_2\!\left(\frac{p_+}{p_-}\right)\right]^2
 =\frac{1-u^2}{4}\!\left[\log_2\!\left(\frac{1+u}{1-u}\right)\right]^2.}
 \label{eq:capacitycopies}
\end{equation}
\rev{For fixed $0<u<1$ and $mC_1\gg1$, the binomial entanglement-energy
profile is asymptotically Gaussian.  The capacity law of
Ref.~\cite{NonlocalMagicSpreading} then gives}
\begin{equation}
 \rev{\mathfrak D_{\rm NL}(\psi_\gamma^{\otimes m})
 \simeq \frac12\log_2\!\left[
 \frac{\pi m(1-u^2)}{32}
 \ln^2\!\left(\frac{1+u}{1-u}\right)\right]
 =\frac12\log_2m+O_\gamma(1).}
 \label{eq:capacityasymptotic}
\end{equation}
\rev{The coefficient $1/2$ does not contradict the coefficient $3/2$ in
Eq.~\eqref{eq:manymagic}: the two equations concern different magic
monotones and different spectral functionals.  They agree only in the
qualitative conclusion that the corresponding regularized density vanishes.
Equation~\eqref{eq:capacityasymptotic} also requires a fixed interior
deformation and cannot be continued uniformly to either endpoint.}

\subsection{\rev{Area-law embedding and entropy-subtracted local magic}}
\rev{The same family gives a direct finite-dimensional specialization of the
entanglement-dominated/magic-dominated distinction of
Ref.~\cite{MagicAreaLaws}.  For every interior deformation, a single pair has
stabilizer group $\{I,Z_AZ_B\}$ of rank one and stabilizer nullity one.
Additivity gives}
\begin{equation}
 \rev{\nu(\psi_\gamma^{\otimes m})=m,\qquad
 S(A^m)=m h_2(p_+),\qquad
 M_\alpha(\psi_\gamma^{\otimes m})=mM_\alpha(\psi_\gamma)=\Theta(m).}
 \label{eq:MDclassification}
\end{equation}
\rev{Hence the fixed interior family satisfies the sufficient extensive-SRE
criterion for the magic-dominated class.  More explicitly, for the specified
cuts $A^m:B^m$ one has
$\min\{|A^m|,|B^m|\}=m=\omega(\log(2m))$ and
$S(A^m)=\Theta(m)=O(\nu)$, which is precisely the MD condition for every
fixed $0<u<1$.  If the copies are embedded as
disjoint short-range dimers on a lattice, a spatial region cuts only dimers
intersecting its boundary and its entropy is
$N_{\partial A}h_2(p_+)$.  This is an area-law realization.  The geometry is
an additional specification: a tensor-copy limit by itself is not a spatial
area law.  Moreover, the state-agnostic hardness theorem of
Ref.~\cite{MagicAreaLaws} is a worst-case uniform statement and does not make
this explicitly known, exactly solvable family computationally hard.}

\rev{Here the one-sided reduced state means the state accessible to Alice
when subsystem $B$ is not observed.  It is the unique operator $\rho_A$ that
reproduces every expectation value of an $A$-local observable,
$\operatorname{Tr}_{AB}[(O_A\otimes I_B)\rho_{AB}]
=\operatorname{Tr}_A(O_A\rho_A)$, and is obtained by the partial trace}
\begin{equation}
 \rev{\rho_A\equiv\operatorname{Tr}_B(\rho_{AB})
 =\sum_{j=0}^{1}{}_B\!\langle j|\rho_{AB}|j\rangle_B.}
 \label{eq:partialTraceDefinition}
\end{equation}
\rev{For the pure right state
$
\vert{}\psi_\gamma\rangle = a\vert{}00\rangle + b\vert{}11\rangle 
= \sqrt{p_+} \vert{}0\rangle_A \vert{}0\rangle_B + \sqrt{p_-} \vert{}1\rangle_A \vert{}1\rangle_B$, with
$a=\sqrt{p_+}$ and $b=\sqrt{p_-}$, the bipartite density operator is}
\begin{align}
 \rev{\rho_{AB}=|\psi_\gamma\rangle\langle\psi_\gamma|}
 &=\rev{p_+|00\rangle\langle00|+\sqrt{p_+p_-}|00\rangle\langle11|}
 \notag\\
 &\quad\rev{+\sqrt{p_+p_-}|11\rangle\langle00|
 +p_-|11\rangle\langle11|.}
 \label{eq:pureBipartiteDensity}
\end{align}
\rev{The identity
$\operatorname{Tr}_B(|i\rangle\langle k|_A\otimes
|j\rangle\langle l|_B)=\delta_{jl}|i\rangle\langle k|_A$
annihilates both cross terms because
$\langle1|0\rangle_B=\langle0|1\rangle_B=0$. $\langle 0\vert{}_B \big( \vert{}0\rangle_A \vert{}0\rangle_B \langle 0\vert{}_A \langle 0\vert{}_B \big) \vert{}0\rangle_B
 = \vert{}0\rangle_A \langle 0\vert{}_A \cdot \langle 0\vert{}0\rangle_B \langle 0\vert{}0\rangle_B = \vert{}0\rangle_A \langle 0\vert{}_A$,
  $\langle 1\vert{}_B \big( \vert{}1\rangle_A \vert{}1\rangle_B \langle 1\vert{}_A \langle 1\vert{}_B \big) \vert{}1\rangle_B 
  = \vert{}1\rangle_A \langle 1\vert{}_A \cdot \langle 1\vert{}1\rangle_B \langle 1\vert{}1\rangle_B = \vert{}1\rangle_A \langle 1\vert{}_A$,
$\langle 0\vert{}_B \big( \vert{}0\rangle_A \vert{}0\rangle_B \langle 1\vert{}_A \langle 1\vert{}_B \big) \vert{}0\rangle_B 
= \vert{}0\rangle_A \langle 1\vert{}_A \cdot \langle 0\vert{}0\rangle_B \langle 1\vert{}0\rangle_B = 0$.
 Consequently,}
\begin{align}
 \rev{\rho_A}
 &\rev{=p_+|0\rangle\langle0|+p_-|1\rangle\langle1|}
 \notag\\
 &\rev{=\begin{pmatrix}p_+&0\\0&p_-\end{pmatrix}
 =\frac12\begin{pmatrix}1+u&0\\0&1-u\end{pmatrix}
 =\frac{I+uZ}{2}.}
 \label{eq:reducedStateDerivation}
\end{align}
\rev{The same reduced state is obtained from the dephased family because
the parameter $t$ changes only the $|00\rangle\langle11|$ coherence, which
vanishes under $\operatorname{Tr}_B$.  Its Pauli expectations are
$\operatorname{Tr}(I\rho_A)=1$, $\operatorname{Tr}(Z\rho_A)=u$, and
$\operatorname{Tr}(X\rho_A)=\operatorname{Tr}(Y\rho_A)=0$.  It therefore
permits a closed calculation of the entropy-subtracted local stabilizer
entropy:}
\begin{align}
 \rev{M_2(\rho_A)}&= -\log_2 \left(\frac{\operatorname{Tr}(\rho_A I)^4 + \operatorname{Tr}(\rho_A Z)^4}{2}\right) \rev{=-\log_2\!\left(\frac{1+u^4}{2}\right),\\
 S_2(\rho_A)&= -\log_2 \operatorname{Tr}(\rho_A^2)=-\log_2\!\left(\frac{1+u^2}{2}\right),}\notag\\
 \rev{\widetilde M_2(\rho_A)}
 &\rev{=M_2(\rho_A)-S_2(\rho_A)
 =\log_2\!\left(\frac{1+u^2}{1+u^4}\right).}
 \label{eq:localReducedMagic}
\end{align}
\rev{It is additive on $\rho_A^{\otimes m}$ and reaches
$\log_2[(1+\sqrt2)/2]\simeq0.271553$ per copy at
$u^2=\sqrt2-1$.  By contrast, the one-copy global nonlocal profile
$\mathcal M_\gamma=-\log_2(1-u^2+u^4)$ is maximal at $u^2=1/2$, where
$\mathcal M_\gamma=\log_2(4/3)$.  The displaced maxima show that tracing out
$B$ and subtracting the R\'enyi-2 mixedness probe a different resource
profile from the globally optimized nonlocal quantity.  In particular,
$\widetilde M_2(\rho_A)$ is a local mixed-state nonstabilizerness diagnostic;
it is neither $\mathcal M_m^{\rm col}$ nor $\mathfrak D_{\rm NL}$, and should
not be interpreted as a universal amount of global magic left after
entanglement has literally been removed.  Indeed,
$\rho_A=p_+|0\rangle\langle0|+p_-|1\rangle\langle1|$ belongs to the convex
stabilizer polytope.  Its nonzero $\widetilde M_2$ diagnoses departure from a
normalized stabilizer-group projector, the free set used for this
entropy-subtracted quantity, rather than distillable magic under the usual
convex-hull stabilizer resource theory.}

\subsection{Finite-copy spectral certificates without a global conjecture}
\label{sec:finiteCB}
Sorted
computational-basis optimality \cite{Sierant} holds at Schmidt rank at most six and gives a
universal gap at larger ranks. The rank condition must be checked for each tensor power.
Let $d=2^m$ and let $\mu_x$ be their Schmidt amplitudes sorted in decreasing
order, including the multiplicities $\binom mj$ of
$\sqrt{p_+^{m-j}p_-^j}$. For the sorted representative
$\sum_x\mu_x\ket{x}_A\ket{x}_B$, an exact Pauli calculation yields
\begin{equation}
 \zeta_{\mathrm{CB}}(m)=\frac1d
 \sum_{a,b\in\mathbb F_2^m}
 \left[\sum_x\mu_x\mu_{x\oplus a}(-1)^{b\cdot x}\right]^4,
 \qquad M_{\mathrm{CB}}(m)=-\log_2\zeta_{\mathrm{CB}}(m).
 \label{eq:Walshpurity}
\end{equation}
Only matched Pauli bit flips on the two sides have nonzero expectation;
the $d$ choices of the redundant phase label produce the prefactor $1/d$.
The inner sum is an unnormalized Walsh transform of
$x\mapsto\mu_x\mu_{x\oplus a}$, so it is evaluated with butterfly additions
without constructing all $4^{2m}$ Pauli matrices.

The factor in Eq.~\eqref{eq:Walshpurity} can be verified from the full
two-party Pauli sum. A Pauli string on either $m$-qubit side has bit-flip
label $a$ and phase label $b$. On the state $\sum_x\mu_x\ket{x,x}$,
the flips must match across sides, since otherwise orthogonality of
the computational labels makes the expectation zero. For matched flips
and phase labels $b_A,b_B$, the magnitude is
\begin{equation}
 \left|\sum_x\mu_x\mu_{x\oplus a}
                 (-1)^{(b_A\oplus b_B)\cdot x}\right|.
 \label{eq:matchedPaulis}
\end{equation}
For each $b=b_A\oplus b_B$, there are $d$ pairs of phase labels. The
overall normalization is $1/d^2$ for $2m$ qubits; their degeneracy therefore
leaves $1/d$. Possible Hermitian Pauli phases have unit magnitude and
drop out of the fourth power. This yields Eq.~\eqref{eq:Walshpurity}
without enumerating a $d^4$ list of dense matrices.

For $m=1,2$, the Schmidt rank is at most four, giving the exact value
$M_2^{\NL}=M_{\mathrm{CB}}$. For arbitrary $m$ the certified interval is
\begin{equation}
 \max\{0,M_{\mathrm{CB}}(m)-4\}
 \le M_2^{\NL}(\psi_\gamma^{\otimes m})\le M_{\mathrm{CB}}(m).
 \label{eq:CBinterval}
\end{equation}
The lower endpoint can be zero at modest copy numbers. For $m\ge3$ we
do not assert exact global optimality merely because the spectrum can be
sorted; its full Schmidt rank exceeds six. Sorting is a collective local
basis permutation, and the resulting Pauli-frame value need not equal
$m\mathcal M_\gamma$. The logarithmic asymptotic law in
Eq.~\eqref{eq:manymagic} remains a separately established theorem.

\begin{figure}[htbp]
\centering
\includegraphics[width=0.75\linewidth]{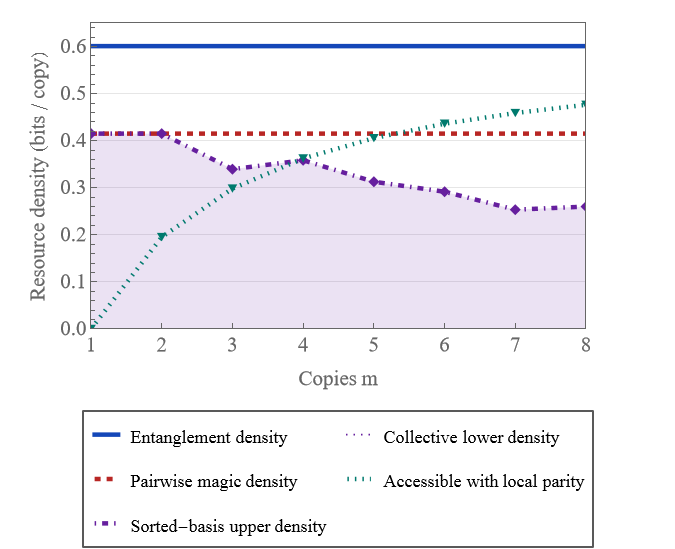}
\caption{Many-copy resource densities at $\gamma=\gamma_*$. The pairwise
magic minimum and ordinary entanglement density are constant. Entanglement
accessible with local parity preserved is given by Eq.~\eqref{eq:parityactivation}
and approaches the ordinary density, despite being zero at one copy.
The sorted-basis magic upper density is computed by the Walsh sum; the shaded
interval is the certified unrestricted range from Eq.~\eqref{eq:CBinterval}.
Its upper endpoint is exact at one and two copies. No asymptotic logarithm
is fitted to finite-copy values.}
\label{fig:spectralcopies}
\end{figure}

\rev{Figure~\ref{fig:spectralcopies} makes the operation-class dependence
explicit.  Ordinary entanglement and the pairwise optimized magic remain
extensive, while parity-restricted accessible entanglement approaches the
ordinary entanglement density as the number of copies grows.  In contrast,
the certified collective-magic density decreases over the displayed range.
The shaded region is a rigorous finite-copy interval rather than a fitted
asymptotic law, and its upper edge is known to be exact only for one and two
copies.}

\section{Exact magic concentration from the original pure pair}
\label{sec:magicconcentration}
\subsection{A fixed Clifford decoding and the order-three invariant}
The same normalized right eigenstate supplies a resource for exact universal
magic concentration \cite{Wu,MagicConcentration}. The decoding, invariant,
and closed six- and eight-copy probabilities extend the original bipartite
framework; the syndrome and Gram calculations below make that specialization
self-contained. This application uses global
qubit stabilizer operations: Clifford gates, stabilizer ancillas, Pauli
measurements, classical feedforward, and discarding. It is different from
local-unitary minimization across a spatial bipartition.

Let $a=\sqrt{p_+}$ and $b=\sqrt{p_-}$ on the positive-phase branch.
Since CNOT is its own inverse,
\begin{equation}
 \operatorname{CNOT}_{A\to B}\ket{\psi_\gamma}
 =\ket{\phi_\gamma}_A\ket0_B,\qquad
 \ket{\phi_\gamma}=a\ket0+b\ket1.
 \label{eq:magicdecoding}
\end{equation}
The decoding is independent of $\gamma$. The input is an unknown pure
qubit state after discarding a stabilizer ancilla, and no state-dependent
rotation is needed. Its Bloch vector is $(c,0,u)$. In particular, it is
non-stabilizer for every fixed $0<|\gamma|<\infty$.

For a pure $n$-qubit state, distinguish the order-three stabilizer purity,
its logarithmic entropy, and its linearized form:
\begin{equation}
 P_3(\varphi)=2^{-n}\sum_{P\in\mathcal P_n}\langle P\rangle_\varphi^6,
 \quad M_3^S(\varphi)=-\tfrac12\log_2P_3(\varphi),\quad
 M_3^{\mathrm{lin}}(\varphi)=1-P_3(\varphi).
 \label{eq:orderthree}
\end{equation}
These are fixed-frame, Clifford-invariant quantities, not an assumed
order-three local-unitary minimum. For the decoded state the four Pauli
expectations are $1,c,0,u$. Thus
\begin{align}
 P_3(\phi_\gamma)&=\tfrac12(1+u^6+c^6)
 =1-\tfrac32u^2c^2,\notag\\
 M_3^{\mathrm{lin}}(\phi_\gamma)&=\tfrac32u^2c^2,\qquad
 M_3^S(\phi_\gamma)=-\tfrac12\log_2(1-\tfrac32u^2c^2).
 \label{eq:orderthreeclosed}
\end{align}
The simplification uses
$u^6+c^6=(u^2+c^2)^3-3u^2c^2(u^2+c^2)=1-3u^2c^2$.
For the original pair, the six nonzero expectations in Eq.~\eqref{eq:SRE}
give $P_3(\psi_\gamma)=\frac14(2+2u^6+2c^6)$, the same value.
This checks Clifford invariance without changing the original state model.

\subsection{\rev{Bell-endpoint magic injection and unitary stabilizer entropy}}
\rev{The Bell endpoint of the same family gives an exact point of contact
with entanglement-enhanced magic injection \cite{MagicInjection}.  At $u=0$,
$\ket{\psi_\gamma}=\ket{\Phi^+}$ up to local phases.  Acting on one side with
an arbitrary unitary $U_A$ prepares its normalized Choi state, and the
unitary stabilizer R\'enyi entropy obeys the exact identity}
\begin{equation}
 \rev{H_\alpha(U_A)=M_\alpha^S\!\left[
 (U_A\otimes I)\ket{\Phi^+}\right].}
 \label{eq:ChoiMagicInjection}
\end{equation}
\rev{Thus a maximally entangled stabilizer input converts the magic resource
of the local unitary into state magic; the associated unitary stabilizer
entropy also lower-bounds its non-Clifford gate cost under the assumptions of
Ref.~\cite{MagicInjection}.  This is an endpoint statement, not a new
concentration protocol.  For a generic finite deformation the Schmidt
spectrum is nonflat and the input already carries nonstabilizerness.  The
exact stabilizer-entangled identity above then no longer applies directly;
the non-stabilizer extension in Ref.~\cite{MagicInjection} is qualitative and
numerical beyond leading asymptotics.  It should therefore not be used to
claim monotonic enhancement throughout the full $\gamma$ curve.  This
operation class also differs from the fixed Clifford decoding below, which
extracts the pre-existing state resource without inserting a new local
non-Clifford unitary.}

\subsection{Six-copy syndrome probability and eight-copy concentration}
To implement the multi-copy magic concentration protocol, we consider six identical copies of the bipartite right eigenstate $\vert{}\psi_\gamma\rangle = a\vert{}00\rangle + b\vert{}11\rangle$. Applying the transversal local Clifford operation $\text{CNOT}_{A \to B}^{\otimes 6} = \bigotimes_{j=1}^6 \text{CNOT}_{A_j \to B_j}$ decouples the state across the bipartition:
   $\text{CNOT}_{A \to B}^{\otimes 6} \vert{}\psi_\gamma\rangle^{\otimes 6} = \vert{}\phi_\gamma\rangle_A^{\otimes 6} \otimes \vert{}0\rangle_B^{\otimes 6}$.
Because Clifford unitaries preserve nonstabilizerness and the auxiliary register $\vert{}0\rangle_B^{\otimes 6}$ is a product stabilizer state carrying zero magic, discarding subsystem $B$ retains the full non-local magic resources in subsystem $A$ without any information loss. The resulting six-copy single-qubit state  $\vert{}\phi_\gamma\rangle_A^{\otimes 6}$ on subsystem $A$ belongs to the totally symmetric subspace $\mathrm{Sym}^6(\mathbb{C}^2)$ serves as the effective non-stabilizer input to the six-copy universal concentration measurement circuit, which is subsequently projected by the joint stabilizer projector $\Pi_{--}^{(6)}$.
The six-copy universal protocol \cite{MagicConcentration} first measures
the commuting Paulis $X^{\otimes6}$ and $Z^{\otimes6}$. Its accepted
syndrome projector is
\begin{equation}
 \Pi_{--}^{(6)}=\tfrac14(I-X^{\otimes6})(I-Z^{\otimes6})
 =\tfrac14(I-X^{\otimes6}-Y^{\otimes6}-Z^{\otimes6}),
 \label{eq:sixsyndrome}
\end{equation}
where the orthogonal projectors (corresponding to eigenvalue of $\mp$1) of $X^{\otimes 6}$ and $Z^{\otimes 6}$ are $\Pi_{\mp}^{(X)} = \frac{I \mp X^{\otimes 6}}{2}$, $\Pi_{\mp}^{(Z)} = \frac{I \mp Z^{\otimes 6}}{2}$, $X^{\otimes 6} Z^{\otimes 6} = (XZ)^{\otimes 6} =(-ZX)^{\otimes 6} = (-i Y)^{\otimes 6} = (-i)^6 Y^{\otimes 6}= - Y^{\otimes 6}$.

To see how the six-copy protocol selectively purifies the nonstabilizer content, 
consider the spectral decomposition of the six-qubit decoded input state 
$|\phi_\gamma\rangle^{\otimes 6} = (a|0\rangle + b|1\rangle)^{\otimes 6}$ with respect to the 
commuting Pauli parity and reflection observables $P \in \{Z^{\otimes 6}, X^{\otimes 6}\}$. 
Any state can be orthogonally resolved into parity eigenspaces as 
$|\phi_\gamma\rangle^{\otimes 6} = |\psi_+\rangle_P + |\psi_-\rangle_P$, where 
$P|\psi_\pm\rangle_P = \pm |\psi_\pm\rangle_P$. The corresponding spectral projectors 
$\Pi_\pm^{(P)} = (I \pm P)/2$ extract these components algebraically via
$\Pi_{\mp}^{(P)} |\phi_\gamma\rangle^{\otimes 6} = |\psi_{\mp}\rangle_P = \mp P |\psi_{\mp}\rangle_P$. 
Filtering onto the joint $-1$ eigenspaces via $\Pi_{--}^{(6)} = \Pi_-^{(X^{\otimes 6})} \Pi_-^{(Z^{\otimes 6})}$ 
is essential: all six single-qubit stabilizer states $s \in Stab_1 = \{\vert{}0\rangle, \vert{}1\rangle, \vert{}+\rangle, \vert{}-\rangle, \vert{}+i\rangle, \vert{}-i\rangle\}$ obey 
$X^{\otimes 6}|s\rangle^{\otimes 6} = Z^{\otimes 6}|s\rangle^{\otimes 6} = +|s\rangle^{\otimes 6}$, 
so projecting onto $(+1,+1)$ would overwhelmingly retain trivial stabilizer contributions. 
Expanding the symmetric state $|\phi_\gamma\rangle^{\otimes 6}$ in the orthonormal Dicke basis 
$|D_w^{(6)}\rangle = \binom{6}{w}^{-1/2} \sum_{|x|=w} |x\rangle$ with Hamming weight $w$ gives $|\phi_\gamma\rangle^{\otimes 6}$ in Eq.\ref{1311}.
The condition $Z^{\otimes 6} = -I$ eliminates all even-weight configurations $w \in \{0,2,4,6\}$ ($\mathcal{H}_+^{(Z)} = \operatorname{span}\{\vert{}D_0^{(6)}\rangle, \vert{}D_2^{(6)}\rangle, \vert{}D_4^{(6)}\rangle, \vert{}D_6^{(6)}\rangle\}$), 
leaving the odd-weight subspace spanned by $\mathcal{H}_-^{(Z)} = \operatorname{span}\{\vert{}D_1^{(6)}\rangle, \vert{}D_3^{(6)}\rangle, \vert{}D_5^{(6)}\rangle\}$.
Subsequent projection by $\Pi_-^{(X^{\otimes 6})}$ exploits the bit-flip symmetry 
$X^{\otimes 6}|D_w^{(6)}\rangle = |D_{6-w}^{(6)}\rangle$, which acts as $+1$ on $|D_3^{(6)}\rangle$ 
and exchanges $|D_1^{(6)}\rangle \leftrightarrow |D_5^{(6)}\rangle$. 
Consequently, the joint $(-1,-1)$ subspace within $Sym^6(\mathbb{C}^2)$ collapses to the unique 
one-dimensional ray spanned by the normalized antisymmetric Dicke superposition 
$|A_6\rangle = (|D_1^{(6)}\rangle - |D_5^{(6)}\rangle)/\sqrt{2}$. 
Evaluating the overlap yields
\begin{equation}
\Pi_{--}^{(6)} |\phi_\gamma\rangle^{\otimes 6} 
= \langle A_6 | \phi_\gamma^{\otimes 6} \rangle |A_6\rangle 
= \frac{\sqrt{6} a^5 b - \sqrt{6} a b^5}{\sqrt{2}} |A_6\rangle 
= \sqrt{3} ab (a^4 - b^4) |A_6\rangle 
= \sqrt{3} ab u |A_6\rangle,
\end{equation}
where $a^2 - b^2 = u$ and $a^2 + b^2 = 1$. The successful syndrome acceptance probability 
is therefore $\|\Pi_{--}^{(6)} |\phi_\gamma\rangle^{\otimes 6}\|^2 = 3 a^2 b^2 u^2 = \frac{3}{4} u^2 c^2 = \frac{1}{2} M_3^{\mathrm{lin}}(\phi_\gamma)$, 
confirming that the $-1$ branches completely filter out the stabilizer background and prepare $|A_6\rangle$ deterministically on success.

The parity projector $\Pi_{-}^{(Z)}$ filters the input state.
For an arbitrary computational-basis state $|x\rangle = |x_1 \dots x_6\rangle$ with Hamming weight $|x| = \sum_{i=1}^6 x_i$, the action of $Z^{\otimes 6}$ yields $Z^{\otimes 6}|x\rangle = (-1)^{|x|}|x\rangle$. Because the Dicke state $|D_w^{(6)}\rangle = \binom{6}{w}^{-1/2} \sum_{|x|=w} |x\rangle$ is an equal-weight superposition of basis states with fixed weight $|x|=w$, it is an exact eigenstate of $Z^{\otimes 6}$ satisfying $Z^{\otimes 6}|D_w^{(6)}\rangle = (-1)^w |D_w^{(6)}\rangle$ with eigenvalue $(-1)^w$. Consequently, the projector onto the negative parity subspace acts on the Dicke basis as
\begin{equation}
\left(\frac{I - Z^{\otimes 6}}{2}\right) |D_w^{(6)}\rangle = \frac{1 - (-1)^w}{2} |D_w^{(6)}\rangle =
\begin{cases}
0, & w \in \{0, 2, 4, 6\}, \\
|D_w^{(6)}\rangle, & w \in \{1, 3, 5\}.
\end{cases}
\end{equation}
Decomposing the six-copy decoded state $|\phi_\gamma\rangle_A^{\otimes 6} = \sum_{w=0}^6 \sqrt{\binom{6}{w}} a^{6-w} b^w |D_w^{(6)}\rangle$ into even- and odd-weight components, $|\phi_\gamma\rangle_A^{\otimes 6} = |\psi_+\rangle_Z + |\psi_-\rangle_Z$, the projector annihilates the even component, $(I - Z^{\otimes 6})|\psi_+\rangle_Z = 0$, while acting as an identity on the odd component, $\frac{1}{2}(I - Z^{\otimes 6})|\psi_-\rangle_Z = |\psi_-\rangle_Z$. Subsequent projection with $(I - X^{\otimes 6})/2$ further isolates the unique antisymmetric superposition within the odd subspace, collapsing the state onto $|A_6\rangle = (|D_1^{(6)}\rangle - |D_5^{(6)}\rangle)/\sqrt{2}$.

The subsequent projection with $\frac{1}{2}(I - X^{\otimes 6})$ exhibits an analogous filtering mechanism based on bit-flip reflection symmetry. Since the single-qubit Pauli $X$ operator flips the computational basis states ($X|0\rangle = |1\rangle, X|1\rangle = |0\rangle$), the transversal operator $X^{\otimes 6}$ maps each state $|x\rangle$ of weight $w$ to a state $|\bar{x}\rangle$ of complementary weight $6 - w$. Linearity over the Dicke states yields $X^{\otimes 6}|D_w^{(6)}\rangle = |D_{6-w}^{(6)}\rangle$. Restricting this action to the surviving three-dimensional odd subspace $\operatorname{span}\{|D_1^{(6)}\rangle, |D_3^{(6)}\rangle, |D_5^{(6)}\rangle\}$, the reflection center $|D_3^{(6)}\rangle$ is an eigenstate with eigenvalue $+1$, satisfying $X^{\otimes 6}|D_3^{(6)}\rangle = |D_3^{(6)}\rangle$, while the complementary pair $\{|D_1^{(6)}\rangle, |D_5^{(6)}\rangle\}$ decomposes into symmetric and antisymmetric superpositions:
\begin{equation}
X^{\otimes 6} \left(\frac{|D_1^{(6)}\rangle \pm |D_5^{(6)}\rangle}{\sqrt{2}}\right) = \pm \left(\frac{|D_1^{(6)}\rangle \pm |D_5^{(6)}\rangle}{\sqrt{2}}\right).
\end{equation}
Therefore, the $+1$ eigenspace of $X^{\otimes 6}$ within the odd sector is two-dimensional, spanned by $|D_3^{(6)}\rangle$ and $(|D_1^{(6)}\rangle + |D_5^{(6)}\rangle)/\sqrt{2}$, both of which are annihilated by $\frac{1}{2}(I - X^{\otimes 6})$. In contrast, the $-1$ eigenspace is strictly one-dimensional, spanned uniquely by the antisymmetric Dicke superposition $|A_6\rangle = (|D_1^{(6)}\rangle - |D_5^{(6)}\rangle)/\sqrt{2}$. The joint projector $\Pi_{--}^{(6)} = \frac{1}{4}(I - X^{\otimes 6})(I - Z^{\otimes 6})$ thus completely filters out all six single-qubit stabilizer product states and isolates the target non-stabilizer ray, collapsing the input state onto $\Pi_{--}^{(6)}|\phi_\gamma\rangle_A^{\otimes 6} = \langle A_6 | \phi_\gamma^{\otimes 6}\rangle |A_6\rangle = \sqrt{3}abu |A_6\rangle$.
Thus the projector onto the negative $X^{\otimes 6}$ eigenspace acts as
\begin{equation}
\left(\frac{I - X^{\otimes 6}}{2}\right)|D_3^{(6)}\rangle = 0, \quad \left(\frac{I - X^{\otimes 6}}{2}\right)\left(\frac{|D_1^{(6)}\rangle + |D_5^{(6)}\rangle}{\sqrt{2}}\right) = 0,
\end{equation}
while annihilating neither component of the unique antisymmetric eigenstate
$|A_6\rangle$
satisfying $\frac{1}{2}(I - X^{\otimes 6})|A_6\rangle = |A_6\rangle$. Thus, the joint projector $\Pi_{--}^{(6)} = \frac{1}{4}(I - X^{\otimes 6})(I - Z^{\otimes 6})$ uniquely isolates $|A_6\rangle$ from the input state $|\phi_\gamma\rangle_A^{\otimes 6}$.

The sign follows from $(XZ)^{\otimes6}=(-i)^6Y^{\otimes6}=-Y^{\otimes6}$.
For a pure qubit with Bloch coordinates $(x,y,z)$,
\begin{equation}
 \langle\Pi_{--}^{(6)}\rangle_{\varphi^{\otimes6}}
 =\tfrac14(1-x^6-y^6-z^6)=\tfrac12M_3^{\mathrm{lin}}(\varphi).
 \label{eq:sixsyndromeprob}
\end{equation}
On the symmetric input subspace the accepted sector is one-dimensional.
To verify this for our real input, use normalized Dicke states
$\ket{D_j^{(k)}}=\binom kj^{-1/2}\sum_{|x|=j}\ket x$ ($\langle D_w^{(k)} \vert{} D_{w'}^{(k)}\rangle = \delta_{w, w'}$). The odd-weight
six-copy states have $j=1,3,5$; $X^{\otimes6}$ exchanges $j$ with $6-j$.
The unique antisymmetric combination is
\begin{equation}
 \ket{A_6}=\frac{\ket{D_1^{(6)}}-\ket{D_5^{(6)}}}{\sqrt2},\quad
 \langle A_6|\phi_\gamma^{\otimes6}\rangle
 =\sqrt3\,ab(a^4-b^4)=\sqrt3\,ab\,u,
 \label{eq:sixDicke}
\end{equation}
where 
\begin{align}
&\vert{}\phi_\gamma^{\otimes 6}\rangle_{A}=\vert{}\phi_\gamma^{\otimes 6}\rangle = (a\vert{}0\rangle + b\vert{}1\rangle)^{\otimes 6} = \sum_{x \in \{0,1\}^6} a^{6 - \vert{}x\vert{}} b^{\vert{}x\vert{}} \vert{}x\rangle  \\
&= \sum_{w=0}^6 \left( \sum_{\vert{}x\vert{}=w} a^{6-\vert{}x\vert{}} b^{\vert{}x\vert{}} \vert{}x\rangle \right) = \sum_{w=0}^6 a^{6-w} b^w \left( \sum_{\vert{}x\vert{}=w} \vert{}x\rangle \right) =\sum_{w=0}^6 \sqrt{\binom{6}{w}} a^{6-w} b^w \vert{}D_w^{(6)}\rangle,
\label{1311}
\end{align}
with the norm square $\left\Vert \sum_{x: \vert{}x\vert{}=w} \vert{}x\rangle \right\Vert^2 = \langle \Sigma_w \vert{} \Sigma_w \rangle = \left(\sum_{x: \vert{}x\vert{}=w} \langle x\vert{}\right) \left(\sum_{y: \vert{}y\vert{}=w} \vert{}y\rangle\right) = \sum_{x: \vert{}x\vert{}=w} \sum_{y: \vert{}y\vert{}=w} \langle x \vert{} y \rangle = \sum_{x: \vert{}x\vert{}=w} \langle x \vert{} x \rangle = \sum_{x: \vert{}x\vert{}=w} 1=\binom{6}{w}$,
$\sum_{\vert{}x\vert{}=w} \vert{}x\rangle = \sqrt{\binom{6}{w}} \vert{}D_w^{(6)}\rangle $, and $\langle D_1^{(6)} \vert{} \phi_\gamma^{\otimes 6}\rangle = \sqrt{\binom{6}{1}} a^{6-1} b^1 = \sqrt{6} a^5 b $, $\langle D_5^{(6)} \vert{} \phi_\gamma^{\otimes 6}\rangle = \sqrt{\binom{6}{5}} a^{6-5} b^5 = \sqrt{6} a b^5$ ($a^2 + b^2 = 1$).
Its squared overlap is $3a^2b^2u^2=3u^2c^2/4$.
The fixed subsequent stabilizer decoding succeeds with probability $2/3$
\cite{MagicConcentration}, so the total six-copy probability is
\begin{equation}
 P_6(\gamma)=\tfrac23\langle\Pi_{--}^{(6)}\rangle
 =\tfrac13M_3^{\mathrm{lin}}(\phi_\gamma)=\tfrac12u_\gamma^2c_\gamma^2.
 \label{eq:CCZsix}
\end{equation}
The output on success is the exact state
$\ket{\mathrm{CCZ}}=\operatorname{CCZ}\ket{+++}$, with the herald and
Clifford corrections retained. Six is the minimal block size within the
class of protocols universal over all pure non-stabilizer qubit inputs;
its success probability is optimal within that universal six-copy class.
This statement does not claim six-copy minimality for a known special
state or for protocols designed only for a restricted input curve.

The eight-copy protocol uses the stabilizer-orthogonal symmetric subspace
$\mathcal V_8$ and accepts a fraction $4/7$ of its input weight
\cite{MagicConcentration}. The weights can be reproduced without a large
tensor calculation. Let $s_r$ run over the six one-qubit Pauli eigenstates,
and define
\begin{equation}
 G^{(k)}_{rs}=\langle s_r|s_s\rangle^k,\quad
 v^{(k)}_r=\langle s_r|\phi_\gamma\rangle^k,\quad
 W_k=1-v^{(k)\dagger}(G^{(k)})^{-1}v^{(k)}.
 \label{eq:stabilizerGramprojection}
\end{equation}
For $k=6,8$ the six stabilizer-copy vectors are linearly independent.
Their span projector is $S_k(G^{(k)})^{-1}S_k^\dagger$, where the columns
of $S_k$ are $\ket{s_r}^{\otimes k}$. Since $\phi_\gamma^{\otimes k}$
is symmetric, subtracting that projector within the symmetric space gives
$W_k=\langle\Pi_{\mathcal V_k}\rangle$. With real $a,b$, direct inversion
of this six-by-six Gram matrix gives the homogeneous identities
\begin{align}
 (a^2+b^2)^6-v^{(6)\dagger}(G^{(6)})^{-1}v^{(6)}
 &=3a^2b^2(a^2-b^2)^2(a^2+b^2)^2,\notag\\
 (a^2+b^2)^8-v^{(8)\dagger}(G^{(8)})^{-1}v^{(8)}
 &=7a^2b^2(a^2-b^2)^2(a^2+b^2)^4.
 \label{eq:Gramweights}
\end{align}
Normalization therefore gives $W_6=3u^2c^2/4$ and $W_8=7u^2c^2/4$.
The eight-copy probability is consequently
\begin{equation}
 P_8(\gamma)=\tfrac47W_8
 =\tfrac23M_3^{\mathrm{lin}}(\phi_\gamma)=u_\gamma^2c_\gamma^2.
 \label{eq:CCZeight}
\end{equation}
Its achievability is established, but optimality at eight copies is not
assumed. Both probabilities peak at $u^2=1/2$, giving
$P_6(\gamma_*)=1/8$ and $P_8(\gamma_*)=1/4$, and vanish at the product
and Bell endpoints where the original pair is a stabilizer state.

\subsection{Direct relation between nonlocal magic and concentration probability}
The order-two purity is
\(2^{-\mathcal M_\gamma}=1-u_\gamma^2c_\gamma^2\). Combining this identity
with Eqs.~\eqref{eq:CCZsix} and \eqref{eq:CCZeight} gives
\begin{align}
 P_6&=\tfrac12(1-2^{-\mathcal M_\gamma}),\notag\\
 P_8&=1-2^{-\mathcal M_\gamma}.
 \label{eq:magicconcentrationbridge}
\end{align}
This is a direct relation between an intrinsic pure-state magic profile and
the success probabilities of a specified exact stabilizer protocol. It does
not make either quantity a complete entanglement ordering. Indeed, for
\(0<P_8<1/4\), the two solutions
\begin{equation}
 u_\pm^2=\frac{1\pm\sqrt{1-4P_8}}2
 \label{eq:concentrationdegeneracy}
\end{equation}
have identical \(P_6,P_8,\mathcal M_\gamma\), but different
\(h_2((1+u_\pm)/2)\). The concentration invariant selects a magic profile
rather than the full Schmidt ordering.

Applying the eight-copy protocol independently on $B$ blocks yields
\begin{equation}
 S_B\sim\operatorname{Bin}(B,P_8),\quad
 P(S_B\ge1)=1-(1-P_8)^B,\quad
 B_\varepsilon=\left\lceil\frac{\ln\varepsilon}{\ln(1-P_8)}\right\rceil.
 \label{eq:CCZblocks}
\end{equation}
The last expression guarantees at least one success with failure at most
$\varepsilon$ when $P_8>0$, at a cost of $8B_\varepsilon$ input pairs.
The law of large numbers gives $S_B/(8B)\to P_8/8$; for any fixed
$0<\delta<P_8$, Hoeffding's bound gives
$P[S_B<B(P_8-\delta)]\le e^{-2B\delta^2}$. Thus the global stabilizer
concentration rate obeys
\begin{equation}
 R_{\mathrm{CCZ}}(\psi_\gamma)\ge\frac{u^2c^2}{8}>0
 \quad(0<|\gamma|<\infty).
 \label{eq:CCZrate}
\end{equation}
This is a lower bound, not the optimized asymptotic rate. The cost is per
original two-qubit pair; it is half as large per input physical qubit.
If a $T$-state catalyst is additionally supplied, deterministic catalytic
conversion of one CCZ resource into two net $T$ states gives the conditional
bound $R_T^{\mathrm{cat}}\ge u^2c^2/4$ \cite{MagicConcentration}.
Such a magic catalyst is distinct from correlated PPT catalysis.
An asymptotic relative-entropy upper bound \cite{MagicConcentration} also
specializes to this same invariant. Writing $L_\gamma=3u^2c^2/2$ gives
\begin{equation}
 \frac{u^2c^2}{8}\le R_{\mathrm{CCZ}}(\psi_\gamma)
 \le\min\!\left\{\frac13,
 C L_\gamma\log_2\frac1{L_\gamma}\right\},\qquad C<1.81.
 \label{eq:CCZratebounds}
\end{equation}
The first upper bound follows separately from stabilizer nullity: each
non-stabilizer input pair has nullity one, each exact CCZ output has nullity
three, nullity adds on products, and cannot increase along a nonzero
stabilizer branch. The second bound controls the weak-resource scaling
up to a logarithmic factor; neither establishes equality with the achievable
eight-copy yield.

The positive rate in Eq.~\eqref{eq:CCZrate} coexists with the zero
collective nonlocal-magic density of Eq.~\eqref{eq:contrast}. Global
stabilizer processing and the global CNOT decoding do not belong to the
local-unitary optimization domain $U(2^m)\otimes U(2^m)$ across separated
laboratories. Conversely unrestricted local unitaries are not generally
stabilizer-free. No contradiction arises between the two statements.

\subsection{Why the pure probabilities do not extend to iid dephased inputs}
The same Clifford decoding maps the mixed family to
\begin{equation}
 \operatorname{CNOT}\rho(u,t)\operatorname{CNOT}^
 \dagger=\omega(u,t)\otimes\ket0\bra0,\qquad
 \omega=\tfrac12\begin{pmatrix}1+u&t\\t&1-u\end{pmatrix}.
 \label{eq:mixedmagicdecoding}
\end{equation}
For $u^2+t^2<1$, $\omega$ is full rank, with minimum eigenvalue
$\lambda_-=(1-\sqrt{u^2+t^2})/2>0$. In particular its iid $k$-copy state
is not supported only on the symmetric subspace. The exact concentration
extension for mixed states confined to that subspace
\cite{MagicConcentration} does not apply to these iid inputs.

The obstruction can be shown directly for any finite stabilizer branch.
Suppose a nonzero accepted CP map $\Lambda$ produced an exact non-stabilizer
pure target $\tau$ from $\omega^{\otimes k}$. Positivity and the spectral
bounds give
\begin{equation}
 \lambda_-^k\Lambda(I)\preceq\Lambda(\omega^{\otimes k})=p\tau
 \preceq\lambda_+^k\Lambda(I),\qquad p>0.
 \label{eq:fullrankobstruction}
\end{equation}
The first inequality forces the support of $\Lambda(I)$ into the
one-dimensional support of $\tau$; the second makes it nonzero. But
$I/2^k$ is a stabilizer mixture, so a stabilizer branch cannot map it to a
non-stabilizer pure state. This is a contradiction. Free ancillas in the
decoded input do not change the argument. Therefore exact finite-block
CCZ concentration has zero success probability for this full-rank logical
input, although approximate distillation is a separate question and qubit
non-stabilizerness can remain present. The pure formulas $P_6,P_8$ must
not be plotted as noisy-state probabilities merely by replacing $c$ by $t$.

\section{Symmetry-resolved entanglement and many-copy activation}
\label{sec:symmetryresolved}
\subsection{Sector weights versus entropy within a sector}
The original local-to-global symmetry reduction can be retained explicitly
\cite{Wu,SymmetryBH}. With $G=X_A X_B$ and $P=Z_AZ_B$,
\begin{equation}
 [H_0,Z_A]=[H_0,Z_B]=[H_0,P]=0,\quad
 \{G,Z_A\}=\{G,Z_B\}=0,\quad [G,P]=0.
 \label{eq:symmetryReductionGenerator}
\end{equation}
It follows that $[S_\gamma,P]=0$ and $[H_\gamma,P]=0$, whereas the
nonzero off-diagonal entries of Eq.~\eqref{eq:Hexplicit} give
$[H_\gamma,Z_A]\ne0$ and $[H_\gamma,Z_B]\ne0$ for finite nonzero
deformation and distinct reference eigenvalues. The two independent local
parities are thus reduced to one globally conserved parity.
In the computational basis,
\begin{equation}
 G=\ket{10}\bra{01}+\ket{01}\bra{10}
   +\ket{11}\bra{00}+\ket{00}\bra{11}.
 \label{eq:exchangeAndPairing}
\end{equation}
The last two terms create or remove a pair, so this is not merely
number-conserving Bose--Hubbard hopping. The resulting sector analysis uses
the algebra of parity blocks, rather than a chaotic Floquet approximation.

Symmetry-resolved reduction \cite{SymmetryBH} separates the weights of
orthogonal local sectors from the states within them. For a block-diagonal
reduced state $\rho_A=\bigoplus_q w_q\rho_{A,q}$, with normalized nonzero
blocks, its purity and von Neumann entropy are
\begin{align}
 \Tr\rho_A^2&=\sum_qw_q^2\Tr\rho_{A,q}^2,\notag\\
 S(\rho_A)&=H(w)+\sum_qw_qS(\rho_{A,q}),\qquad
 H(w)=-\sum_qw_q\log_2w_q.
 \label{eq:sectoridentities}
\end{align}
The logarithm acts within each block as
$\log(w_q\rho_{A,q})=(\log w_q)I+\log\rho_{A,q}$, proving the second
identity; orthogonal supports eliminate cross terms in the first.
For the linear entropy used in symmetry-resolved random-matrix calculations
\cite{SymmetryBH}, the corresponding identity is
\begin{equation}
 S_{\mathrm{lin}}=S_{\mathrm{num,lin}}+S_{\mathrm{conf,lin}},\quad
 S_{\mathrm{num,lin}}=1-\sum_qw_q^2,\quad
 S_{\mathrm{conf,lin}}=\sum_qw_q^2(1-\Tr\rho_{A,q}^2).
 \label{eq:linearSector}
\end{equation}
Its configurational weights are $w_q^2$, whereas the von Neumann version
uses $w_q$. Linear entropy, logarithmic R\'enyi entropy, and stabilizer
R\'enyi entropy are consequently not interchangeable.

The original state has fixed global $Z_2$ parity, rather than fixed total
particle number: its $\ket{00}$ and $\ket{11}$ components have occupation
numbers zero and two. Its local parity sectors are one-dimensional, so
\begin{equation}
 w_+=p_+,\quad w_-=p_-,\quad
 S_{\mathrm{conf,lin}}=0,\quad S_{\mathrm{lin}}=c^2/2,\quad
 S(\rho_A)=h_2(p_+).
 \label{eq:singleSector}
\end{equation}
This is an algebraic decomposition of the original entropy. Under the
ordinary qubit LOCC of Sec.~\ref{sec:distillation}, that entropy remains
distillable entanglement; being carried by sector weights does not make it
non-quantum. A different operational conclusion requires imposing a local
superselection restriction, which is an additional assumption here.

\subsection{Sector uncertainty is not a mixed-state entanglement measure}
\label{sec:sectorMeaning}
There are three distinct entropies of the sector weights. For this pair,
\begin{align}
 H_{\mathrm{sec}}&=-p_+\log_2p_+-p_-\log_2p_-=h_2(p_+),\notag\\
 H_{\mathrm{sec}}^{(2)}&=-\log_2(p_+^2+p_-^2)
                         =-\log_2\frac{1+u^2}{2},\notag\\
 S_{\mathrm{num,lin}}&=1-p_+^2-p_-^2=\frac{1-u^2}{2}=\frac{c^2}{2}.
 \label{eq:threeSectorEntropies}
\end{align}
The first is Shannon number entropy when the labels are particle numbers,
or Shannon sector entropy for general symmetry labels. The second is
order-two R\'enyi number entropy. The last is the linear, or quadratic,
number-entropy contribution of Eq.~\eqref{eq:linearSector}; it is not
itself a Shannon entropy. All three are functions of sector probabilities,
but their normalizations, additivity properties, and operational roles differ.

For the present $Z_2$ sectors,
$\rho_{A,+}=\ket0\bra0$ and $\rho_{A,-}=\ket1\bra1$.
Consequently each normalized block has zero entropy and unit purity, so
$S_{\mathrm{conf,lin}}=0$. For the pure joint state,
$1-\Tr\rho_A^2=S_{\mathrm{num,lin}}$ is nevertheless a genuine
bipartite pure-state entanglement measure. The coherent amplitudes
$a\ket{00}+b\ket{11}$ produce this entanglement; its numerical equality
to a classical entropy of the sector label does not make the joint state
a classical mixture. The sector decomposition identifies where the
Schmidt uncertainty resides, not a classical mechanism generating entanglement.

This distinction is tested by local dephasing at fixed marginal $u$.
For $0\le\eta\le1$, define
\begin{equation}
 \rho_\eta=\rho(u,\eta c)
 =p_+\ket{00}\bra{00}+p_-\ket{11}\bra{11}
 +\eta\sqrt{p_+p_-}
       (\ket{00}\bra{11}+\ket{11}\bra{00}).
 \label{eq:sectorCoherenceFamily}
\end{equation}
Partial tracing removes the cross terms for every $\eta$, giving
\begin{equation}
 (\rho_\eta)_A=(\rho_\eta)_B
 =p_+\ket0\bra0+p_-\ket1\bra1.
 \label{eq:sectorFixedMarginals}
\end{equation}
Hence every entropy in Eq.~\eqref{eq:threeSectorEntropies}, and every
one-dimensional configurational contribution, stays unchanged. In contrast,
\begin{align}
 \operatorname{spec}(\rho_\eta^{T_B})
 &=\{p_+,p_-,+\eta c/2,-\eta c/2\},\notag\\
 \mathcal N(\rho_\eta)&=\eta c/2,\qquad
 \Tr\rho_\eta^2=\tfrac12(1+u^2+\eta^2c^2),\notag\\
 E_d(\rho_\eta)&=h_2(p_+)
 -h_2\!\left(\frac{1+\sqrt{u^2+\eta^2c^2}}2\right).
 \label{eq:sectorSameWeightsDifferentResources}
\end{align}
The partial transpose has the two populations as diagonal eigenvalues and
a separate two-by-two block with off-diagonal entry $\eta c/2$, giving
the displayed positive and negative pair. The joint purity follows by
squaring its original support block. These are direct coherence-sensitive
tests of the difference that the reduced sector distribution cannot detect.

At $\eta=0$ and $0\le u<1$, the joint state is separable but
\begin{equation}
 S_{\mathrm{num,lin}}=\frac{1-u^2}{2}>0,\qquad
 H_{\mathrm{sec}}=h_2(p_+)>0,\qquad
 E_d=\mathcal N=0.
 \label{eq:separablePositiveNumberEntropy}
\end{equation}
For example $u=0$ gives the separable mixture
$(\ket{00}\bra{00}+\ket{11}\bra{11})/2$, with linear number entropy
$1/2$ and Shannon number entropy one bit. At the pure product endpoint
$u=1$, the weights instead become $(1,0)$ and all three sector entropies
vanish. Separability alone does not force deterministic sector weights.

Insensitivity to within-sector spatial correlations means insensitivity
at fixed sector weights: changing $\rho_{A,q}$ while keeping $w_q$
cannot change $H(w)$ or $1-\sum_qw_q^2$. The weights can themselves
depend on correlations in a general state, so the statement is not
independence from every possible change of spatial correlations. Here each
single-pair sector is one-dimensional, and there are no within-sector
degrees of freedom to correlate. Entanglement is defined across the fixed
physical $A:B$ tensor product; when the qubits are held in separated
regions this is spatial bipartite entanglement, without implying an extended
lattice or a thermodynamic limit.

\subsection{Local parity restriction and activation on several copies}
Suppose each laboratory must preserve its total local parity, with no shared
parity reference. For a pure state of fixed global parity, project the local
sectors, retain the common sector label, and allow arbitrary operations within
each sector. The accessible entanglement of this pure ensemble is
$E_{\mathrm{acc},Z_2}=\sum_qw_qS(\rho_{A,q})$. On a single pair it is
zero: local parity twirling removes the $\ket{00}\bra{11}$ coherence,
leaving the separable comparison state in Eq.~\eqref{eq:sigma}.

For $m$ original pairs, let $j$ count occupied modes on one side. The local
parity sector weights follow from the binomial spectrum:
\begin{align}
 w_{\mathrm e}^{(m)}&=\sum_{j\ \mathrm{even}}\binom mjp_+^{m-j}p_-^j
 =\frac{(p_++p_-)^m+(p_+-p_-)^m}{2}=\frac{1+u^m}{2},\notag\\
 w_{\mathrm o}^{(m)}&=\frac{1-u^m}{2}.
 \label{eq:manyparityweights}
\end{align}
The total reduced entropy is $m h_2(p_+)$. Applying
Eq.~\eqref{eq:sectoridentities} gives the exact finite-copy expression
\begin{equation}
 E_{\mathrm{acc},Z_2}^{(m)}
 =m h_2(p_+)-h_2\!\left(\frac{1+u^m}{2}\right).
 \label{eq:parityactivation}
\end{equation}
For two copies, the even sector has Schmidt probabilities
$p_+^2/(p_+^2+p_-^2)$ and $p_-^2/(p_+^2+p_-^2)$, while the odd sector
is an encoded Bell pair on the local basis $\{\ket{01},\ket{10}\}$.
Explicitly,
\begin{equation}
 E_{\mathrm{acc},Z_2}^{(2)}
 =(p_+^2+p_-^2)h_2\!\left(\frac{p_+^2}{p_+^2+p_-^2}\right)
 +2p_+p_-.
 \label{eq:twoParityActivation}
\end{equation}
This is positive in the interior even though the one-copy accessible value
is zero. The restriction costs at most one bit for every block:
\begin{equation}
 0\le m h_2(p_+)-E_{\mathrm{acc},Z_2}^{(m)}\le1,\qquad
 \lim_{m\to\infty}\frac{E_{\mathrm{acc},Z_2}^{(m)}}m=h_2(p_+).
 \label{eq:paritydensity}
\end{equation}
Thus asymptotically positive accessible entanglement coexists with zero
collective nonlocal-magic density even under this parity restriction.
The restriction is not retroactively imposed on the preceding qubit protocols.

The same calculation in the linear-entropy convention gives
\begin{align}
 S_{\mathrm{num,lin}}^{(m)}&=\frac{1-u^{2m}}2,\notag\\
 S_{\mathrm{conf,lin}}^{(m)}&=\frac{1+u^{2m}}2
                       -\left(\frac{1+u^2}{2}\right)^m,\notag\\
 S_{\mathrm{lin}}^{(m)}&=1-\left(\frac{1+u^2}{2}\right)^m.
 \label{eq:linearParityCopies}
\end{align}
These follow from $w_{\mathrm e}^2+w_{\mathrm o}^2=(1+u^{2m})/2$
and tensor-product purity. They are dimensionless entropies, not rates in
bits. The block formulas here concern pure inputs; the unchanged marginal
entropy of the mixed dephased state is not its distillable entanglement.

\subsection{A stronger occupation-number restriction}
If local operations are additionally required to conserve total occupation
number, the sectors are the full values $j$, not just their parity. This
is a separate $U(1)$ operational restriction, not a claim that
$H_\gamma$ conserves occupation number. The probability of sector $j$ is
\begin{equation}
 b_j=\binom mjp_+^{m-j}p_-^j,
 \qquad E_j=\log_2\binom mj.
 \label{eq:numberSectorCopies}
\end{equation}
Within a fixed $j$, all Schmidt amplitudes are equal, giving a maximally
entangled conditional state of rank $\binom mj$. The entropy identity now is
\begin{equation}
 E_{\mathrm{acc},U(1)}^{(m)}=\sum_jb_j\log_2\binom mj
 =m h_2(p_+)-H[\operatorname{Bin}(m,p_-)].
 \label{eq:numberactivation}
\end{equation}
Expanding $\log_2 b_j=\log_2\binom mj+(m-j)\log_2p_++j\log_2p_-$
and using $\mathbb E j=mp_-$ proves the second equality.
For two copies this reduces to the transparent finite-sector result
\begin{equation}
 E_{\mathrm{acc},U(1)}^{(2)}=2p_+p_- =\frac{c^2}{2}.
 \label{eq:twoCopyNumberExample}
\end{equation}
Combining it with the parity-restricted value in
Eq.~\eqref{eq:twoParityActivation} gives
\begin{equation}
 E_{\mathrm{acc},Z_2}^{(2)}-E_{\mathrm{acc},U(1)}^{(2)}
 =(p_+^2+p_-^2)h_2\!\left(\frac{p_+^2}{p_+^2+p_-^2}\right)>0
 \quad(p_+p_->0).
 \label{eq:twoCopyRestrictionGap}
\end{equation}
Thus a local parity restriction preserves coherence between occupations zero
and two inside the even sector, whereas the stronger occupation-number
restriction removes that contribution.
For fixed $0<p_-<1/2$, Stirling's expansion and the local central limit
approximation yield
\begin{equation}
 H[\operatorname{Bin}(m,p_-)]=\tfrac12\log_2(2\pi e m p_+p_-)+o(1),
 \quad E_{\mathrm{acc},U(1)}^{(m)}=m h_2(p_+)-\tfrac12\log_2m+O_\gamma(1).
 \label{eq:numberasymptotics}
\end{equation}
The Shannon entropy of a width-$\sqrt m$ binomial is logarithmic with
coefficient $1/2$; the shell fourth-moment magic in
Eq.~\eqref{eq:manymagic} has coefficient $3/2$. Both originate in the
same spectrum but measure different functions of its weights.

\section{Parity protection, Gaussian filtering, and fermionic code distance}
\label{sec:fermionicQEC}
\subsection{The covariance matrix and the two-mode parity code}
Fermionic Gaussian error-correction constraints \cite{FermionicQEC} become
transparent when the original occupation-space interpretation is retained.
Use the Jordan--Wigner Majoranas
\begin{equation}
 \xi_1=X\otimes I,\quad\xi_2=Y\otimes I,\quad
 \xi_3=Z\otimes X,\quad\xi_4=Z\otimes Y,\qquad
 \Gamma_{jk}=-\frac i2\Tr(\rho[\xi_j,\xi_k]).
 \label{eq:codeMajoranas}
\end{equation}
Evaluation using Eq.~\eqref{eq:rhoPauli} gives
\begin{equation}
 \Gamma(u,t)=\begin{pmatrix}
 0&u&0&t\\-u&0&t&0\\0&-t&0&u\\-t&0&-u&0
 \end{pmatrix},\qquad\Gamma^2=-(u^2+t^2)I_4.
 \label{eq:codeCovariance}
\end{equation}
For example $-i\xi_1\xi_2=ZI$, $-i\xi_3\xi_4=IZ$,
$-i\xi_1\xi_4=-YY$, and $-i\xi_2\xi_3=XX$.
\rev{The last identity classifies the original similarity transformation
within the fermionic operator language. Namely,}
\begin{equation}
 \rev{\begin{aligned}
 S_\gamma&=e^{\gamma X\otimes X}=e^{-i\gamma\xi_2\xi_3},\\
 S_\gamma\xi_2S_\gamma^{-1}
 &=\cosh(2\gamma)\xi_2+i\sinh(2\gamma)\xi_3,\\
 S_\gamma\xi_3S_\gamma^{-1}
 &=\cosh(2\gamma)\xi_3-i\sinh(2\gamma)\xi_2.
 \end{aligned}}
 \label{eq:complexMajoranaRotation}
\end{equation}
\rev{This is the hyperbolic analytic continuation of a Majorana-bilinear
rotation: it is an invertible nonunitary Gaussian/Bogoliubov transformation
in the complexified fermionic sense, not a physical unitary fermionic
Clifford gate for real nonzero \(\gamma\). The distinction is consistent
with the separate Pauli, Majorana, and Dirac-fermion Clifford classifications
of Ref.~\cite{MagoulasEvangelista2026}.}
On the pure circle $\Gamma^2=-I$, consistent with Gaussianity. Its
Pfaffian is $u^2+t^2$, whereas $\langle Z_AZ_B\rangle=1$.
The mismatch of that quartic expectation with its Wick prediction is
$1-u^2-t^2=4\Delta_{\mathrm{Wick}}$. This recovers the earlier defect
through the Majorana covariance rather than a separate resource definition.

Let $P=\Pi_+$ project onto $\mathcal C=\operatorname{span}\{\ket{00},\ket{11}\}$.
The Majorana distance of a rank-greater-than-one code is
\begin{equation}
 d_F(P)=\min\{|S|:\ P\xi_S P\notin\mathbb C P,\ S\ne\varnothing\}.
 \label{eq:MajoranaDistance}
\end{equation}
Every single Majorana anticommutes with total parity, so $P\xi_jP=0$.
But the weight-two operator $-i\xi_1\xi_2=Z_A$ acts within the code as
the non-scalar logical $Z_L$:
\begin{equation}
 P Z_A P=\ket{00}\bra{00}-\ket{11}\bra{11}\notin\mathbb C P,
 \qquad d_F(P)=2.
 \label{eq:parityCodeDistance}
\end{equation}
Thus it is a two-mode, one-logical-qubit distance-two fermionic code.
Every pure state in this two-dimensional even space is a BCS Gaussian
state; it is not a code of distance at least three.

The Knill--Laflamme criterion for an error set $\{E_a\}$ is
$PE_a^\dagger E_bP=\alpha_{ab}P$. For the restricted noise set
$\{I,Z_AZ_B\}$ every product is scalar on the code, proving its
decoherence-free protection. In contrast $PZ_BP=Z_L$ is non-scalar,
so the same code cannot even detect that local phase error without accessing
logical information. Arbitrary single-Majorana errors are also not
correctable: although each individually moves the state to the opposite
parity sector, their cross products contain non-scalar weight-two logical
operators. A noise-fixed eigenstate and an error-correcting code for arbitrary
logical superpositions are different objects.

\subsection{\rev{Total-parity and six-copy syndrome twirls}}
\rev{For the present two-mode interpretation let
$Q=Z_AZ_B$ and $\Pi_\pm=(I\pm Q)/2$.  Since
$Z_j=1-2n_j=(-1)^{n_j}$,}
\begin{equation}
 \rev{Q=(-1)^{n_A+n_B}=e^{i\pi(n_A+n_B)},\qquad
 \mathcal H_+=\operatorname{span}\{|00\rangle,|11\rangle\},\quad
 \mathcal H_-=\operatorname{span}\{|01\rangle,|10\rangle\}.}
 \label{eq:totalParitySectors}
\end{equation}
\rev{The total-parity twirl is the group average over the two-element group
$\mathbb Z_2=\{I,Q\}$:}
\begin{equation}
 \rev{\mathcal T_Q(\rho)=\frac12(\rho+Q\rho Q)
 =\Pi_+\rho\Pi_++\Pi_-\rho\Pi_-.}
 \label{eq:totalParityTwirl}
\end{equation}
\rev{Operationally, Eq.~\eqref{eq:totalParityTwirl} means applying $I$ with
probability $1/2$ and $Q$ with probability $1/2$, then forgetting which was
applied.  In the parity-ordered basis every density matrix has the block form}
\begin{equation}
 \rev{\rho=\begin{pmatrix}\rho_{++}&\rho_{+-}\\
 \rho_{-+}&\rho_{--}\end{pmatrix},\qquad
 \mathcal T_Q(\rho)=
 \begin{pmatrix}\rho_{++}&0\\0&\rho_{--}\end{pmatrix}.}
 \label{eq:parityBlockDephasing}
\end{equation}
\rev{Indeed, $Q\Pi_\pm=\pm\Pi_\pm$, so conjugation by $Q$ preserves the two
diagonal blocks and reverses the sign of the two off-diagonal blocks.  The
average therefore removes only coherence between the even and odd sectors;
it preserves all structure and population within each sector.}

\rev{Because $n_A+n_B$ is quadratic in fermionic operators, $Q$ is a
Gaussian unitary.  Accordingly, $\mathcal T_Q$ is a random
Gaussian-unitary channel in an operation class that permits classical
randomness.  This statement does not make the projective measurement
$\{\Pi_+,\Pi_-\}$ a free Gaussian instrument, nor does a convex mixture of
Gaussian unitaries automatically satisfy every stricter definition of a
deterministic fermionic Gaussian channel.  In particular, the twirl does not
remove the odd population
$\ell=\operatorname{Tr}(\Pi_-\rho)$ or correct logical errors inside either
sector.  If an even target has preselection fidelity $F$, an additional
even-parity postselection produces}
\begin{equation}
 \rev{p_+=1-\ell,\qquad
 \rho_+=\frac{\Pi_+\rho\Pi_+}{1-\ell},\qquad
 F_+=\frac{F}{1-\ell}.}
 \label{eq:parityPostselection}
\end{equation}
\rev{This conditioning is a distinct, non-trace-preserving step.  Under a
fermion-parity superselection rule, physical states already obey
$[\rho,Q]=0$ and hence $\rho_{+-}=\rho_{-+}=0$; in that case
$\mathcal T_Q(\rho)=\rho$.}

\rev{For the six-copy qubit concentration circuit, averaging over the two
commuting measured Paulis gives instead}
\begin{equation}
 \rev{\mathcal T_6(\rho)=\frac14\sum_{r,s=0}^{1}
 (X^{\otimes6})^r(Z^{\otimes6})^s\rho
 (Z^{\otimes6})^s(X^{\otimes6})^r
 =\sum_{\mu,\nu=\pm}\Pi_{\mu\nu}\rho\Pi_{\mu\nu}.}
 \label{eq:sixCopyTwirl}
\end{equation}
\rev{This stabilizer-unitary twirl preserves every syndrome probability,
including $\operatorname{tr}(\Pi_{--}^{(6)}\rho)$, while eliminating
coherence between the four syndrome sectors.  It is not automatically a
fermionic Gaussian channel.  It concerns the joint eigenvalues of
$X^{\otimes6}$ and $Z^{\otimes6}$, whereas
Eq.~\eqref{eq:totalParityTwirl} concerns the two-mode total fermion parity
$Q$.  Keeping these labels separate prevents a qubit Pauli-syndrome
measurement from being silently promoted to a free nondestructive
fermionic-parity measurement.}

\rev{The qubit CNOT decoder in Eq.~\eqref{eq:magicdecoding}, while
Pauli--Clifford, is not a parity-preserving fermionic Gaussian operation,
because $[\mathrm{CNOT}_{A\to B},Z_AZ_B]\ne0$
\cite{MagoulasEvangelista2026}.}

\subsection{Gaussian postselection can nevertheless concentrate this pair}
The distinction between deterministic Gaussian channels and heralded
Gaussian instruments \cite{FermionicQEC} supplies a direct application to
the existing filter in Eq.~\eqref{eq:filter}. For $m$ copies of the pure
pair, the cross-covariance block has norm $c$, independent of $m$.
A product of local trace-preserving Gaussian channels maps it to
$C_{\mathrm{out}}=M_A C M_B^{\mathsf T}$, with
$\|M_A\|_\infty,\|M_B\|_\infty\le1$, so
\begin{equation}
 \|C_{\mathrm{out}}\|_\infty\le c,\qquad
 F_{\mathrm{EPR}}(\rho_{\mathrm{out}})
 =\tfrac14[1+(C_{\mathrm{out}})_{12}+(C_{\mathrm{out}})_{21}
                  +\langle Z_AZ_B\rangle]\le\frac{1+c}{2}.
 \label{eq:GaussianTPbound}
\end{equation}
The last inequality uses $\langle Z_AZ_B\rangle\le1$ and bounds each
cross entry by the matrix norm. Convex mixtures of such product channels
obey the same bound because fidelity is linear. The input pair already
attains $(1+c)/2$. This bound does not cover postselection and conditional
classical feedforward.

Choose $\lambda=\sqrt{p_-/p_+}$ and the two Kraus operators
\begin{equation}
 K_\lambda=\lambda\ket0\bra0+\ket1\bra1,\qquad
 L_\lambda=\sqrt{1-\lambda^2}\ket1\bra0,
 \quad K_\lambda^\dagger K_\lambda+L_\lambda^\dagger L_\lambda=I.
 \label{eq:GaussianFilterInstrument}
\end{equation}
The success operator is precisely the existing diagonal filter, with
$K_\lambda=\lambda\exp[-(\ln\lambda)n_A]$ when $\lambda>0$.
It is a positive Gaussian operator. It can be implemented by appending an
occupied ancillary mode, applying a number-conserving two-mode rotation
with $\cos\alpha=\lambda$, and retaining the occupied-ancilla outcome;
the other outcome implements $L_\lambda$ up to a phase. Thus both
fine-grained branches are Gaussian.
Their action on the original pair is
\begin{align}
 (K_\lambda\otimes I)\ket{\psi_\gamma}
 &=\sqrt{p_-}(\ket{00}+\ket{11}),\notag\\
 (L_\lambda\otimes I)\ket{\psi_\gamma}
 &=\sqrt{p_+-p_-}\ket{10}=\sqrt u\ket{10}.
 \label{eq:GaussianFilterBranches}
\end{align}
The success probability is $2p_-=1-u$, and its conditional state is an
exact EPR pair. Its conditional fidelity has increased to one, whereas
the unconditioned fidelity is $1-u\le(1+c)/2$, consistent with
Eq.~\eqref{eq:GaussianTPbound}. Alice must communicate the herald to Bob.
For a dephased input the success probability is still $1-u$, but the
conditional output is $\rho(0,\eta)$ with EPR fidelity $(1+\eta)/2$,
not a pure Bell state.

Finally, Gaussian states are not a convex set \cite{FermionicQEC}.
The local phase channel is a classical mixture of Gaussian phase unitaries,
yet its mixed output can have the positive Wick defect of
Eq.~\eqref{eq:Wickdefect}. That defect and the matched-Gaussian relative
entropy quantify a failure of Wick factorization; they are not automatically
monotones under operation classes allowing arbitrary classical Gaussian
mixing. The Gaussian-channel fidelity bound is applied here only to its
specified pure Gaussian input and stated trace-preserving operation class.

\section{Discussion}
The fixed-spectrum deformation furnishes an analytical laboratory in which
entanglement manipulation, nonlocal magic, Gaussian structure, and state
sharing can be compared without conflating eigenvalues with physical
right-state resources. The optimized CHSH value tracks the survival of
coherence in the noisy family, while the exact distillation rate distinguishes
usable entanglement from state fidelity and from classical correlations at the
fully dephased boundary.

The general maximally correlated distillation identity, PPT catalytic
monotonicity, and the imperfect-dimer scaling are established results. The
contributions derived here are their explicit specialization to the deformed
preparation family, the exact noisy-state rate and confidence bound, the
closed many-copy magic profiles, and the direct relation between nonlocal
magic and exact CCZ concentration probability. We do not assert that every
consequence of this specialization constitutes an independent general
resource-theory theorem.

The hashing rate matches every PPT comparison
distance for every tensor power, and the catalytic monotone then gives the
same achievable rate even with correlated catalysts. Secret-key extraction
after a specified measurement and probabilistic local filtering give further
operational comparisons, with the latter generally sacrificing rate.
The finite-sample bound is useful when the family and measurements are
calibrated, while universal unknown-state protocols and black-box gate
certification address broader problems with additional assumptions.

The same per-copy limiting operation has a different outcome for magic.
The original Schmidt-frame entropy is additive, and independently optimized
pairwise local rotations retain that extensive value. Across the same
$A^m:B^m$ cut, collective local rotations preserve the entire entanglement
spectrum but change its assignment to Pauli-frame labels. The binomial
spectrum spreads its shell weight across order $\sqrt m$ shells, whose
fourth-moment participation scales as $m^{-3/2}$. Its logarithm yields
the subextensive collective nonlocal magic, so dividing by $m$ gives zero.
Entanglement density remains $h_2(p_+)>0$. Thus the distinction concerns
which local basis changes are permitted, not the removal of entanglement or
an inferred magic-distillation protocol. The fixed global Clifford decoding
provides a separate positive CCZ yield, $R_{\mathrm{CCZ}}\ge u^2c^2/8$.
This operation class differs from collective local-unitary minimization.
With an added local parity restriction, accessible entanglement loses at
most one bit per block and its density still tends to $h_2(p_+)$.

\rev{The added comparisons sharpen rather than collapse these distinctions.
Stabilizer-fidelity nonlocal magic grows as
$\frac12\log_2m+O_\gamma(1)$, whereas the collective order-two quantity above
grows as $\frac32\log_2m+O_\gamma(1)$; both have zero density because they are
different spectral monotones.  The reduced-state quantity
$\widetilde M_2(\rho_A)$ is local and extensive on copies, while the
short-range dimer embedding realizes its spatial area-law setting.  Random
parity twirling removes sector coherence but not leakage, and the Choi-state
magic-injection identity is exact at the Bell endpoint only.  These results
show explicitly that ``magic,'' its locality, and its asymptotic coefficient
remain operation- and definition-dependent even when every expression is
controlled by the same one-parameter Schmidt spectrum.}

Extensions beyond the maximally correlated support would require new bounds,
as population leakage can destroy the exact hashing/relative-entropy equality.

The resource comparisons also distinguish quantities beyond entanglement
rate. The pure BCS state has zero fermionic non-Gaussianity yet nonzero
stabilizer magic, and local dephasing produces a positive Wick defect.
The fixed-state mechanism is transparent in two symmetry representations:
parity conjugation fixes the physical even-sector density matrix, while
the four-replica reflection fixes three orthogonal difference-mode creation
operators. Original replica modes still mix, and the adjoint fixed algebra
also contains products of sign-flipped modes. Identical BCS inputs are
invariant under any real orthogonal copy rotation through preservation of
their pairing bilinear, a stronger fixed-state statement than the existence
of individual invariant modes. Correlated parity kicks fix the entire
original noisy family but can mix cross-coupled dimers. Target-matched
swapping preserves Schmidt vectors while
powering the weights and generally reducing conditional entropy; a
flat-spectrum auxiliary link restores exact deterministic sharing with
different resource costs. These are finite, controlled applications;
they do not infer CFT, SW-SSB, or SYK thermodynamic phases.
The parity support has Majorana distance two, and its protection against
joint phase kicks does not correct local phase noise. A Gaussian heralded filter can concentrate the original pair into an
EPR state. The difference between channel averaging, retained heralds,
and protected logical information is essential to these conclusions.

\appendix

\section{Preparation circuit and what qubit certification can establish}
\label{app:gate}
The ideal state can be prepared with
\begin{equation}
 \theta_\gamma=\arctan(\tanh|\gamma|),\qquad
 U_{\mathrm{prep}}(\gamma)=\operatorname{CNOT}_{A\to B}
 [R_y(2\theta_\gamma)\otimes I].
 \label{eq:prep}
\end{equation}
The original nonunitary \(S_\gamma\) defines the eigenstate family; the circuit
in Eq.~\eqref{eq:prep} is a unitary preparation of a selected normalized state,
not a unitary implementation of the similarity transformation.

A useful calibrated noise model applies an isotropic depolarizing channel
\(\mathcal D_r(\omega)=r\omega+(1-r)I/2\), \(0\le r\le1\), after the rotation
and before an ideal CNOT. The rotation channel's Pauli-transfer matrix has
eigenvalues
\begin{equation}
 \{1,r,re^{i2\theta},re^{-i2\theta}\}.
 \label{eq:PTM}
\end{equation}
Starting from \(\ket{00}\), the final state is exactly Eq.~\eqref{eq:rho} with
\begin{equation}
 u=r\cos2\theta,\qquad t=r\sin2\theta,\qquad
 \Dcal=\hbin\!\left(\frac{1+r\cos2\theta}{2}\right)
                   -\hbin\!\left(\frac{1+r}{2}\right),
 \label{eq:gaterate}
\end{equation}
for \(0\le\theta\le\pi/4\). This provides a conditional map from calibrated
rotation contraction and phase to an entanglement rate. It is different from
the fixed-population dephasing path in Eq.~\eqref{eq:dephasing}.

The calculation is visible at the one-qubit level. The ideal rotation gives
Bloch vector $(\sin2\theta,0,\cos2\theta)$, and depolarization multiplies
it by $r$. Before the CNOT the control's density matrix is
\begin{equation}
 \omega_A=\frac12\begin{pmatrix}
 1+r\cos2\theta&r\sin2\theta\\r\sin2\theta&1-r\cos2\theta
 \end{pmatrix}.
 \label{eq:controlpreparation}
\end{equation}
The CNOT maps $\ket0\ket0$ to $\ket{00}$ and $\ket1\ket0$ to
$\ket{11}$, transporting this block to the even two-qubit support without
changing its eigenvalues. This proves Eq.~\eqref{eq:gaterate}.
The Pauli-transfer matrix of the rotation followed by this noise is
$1\oplus r\mathsf R_y(2\theta)$; its three-dimensional rotation block
has eigenvalues $1,e^{i2\theta},e^{-i2\theta}$, proving
Eq.~\eqref{eq:PTM}. When the axis and target angle $\theta_0$ are calibrated,
the average gate fidelity has the additional closed form
\begin{equation}
 F_{\mathrm{avg}}(\theta_0;r,\theta)
 =\frac{3+\Tr[\mathsf R_y(2\theta_0)^{\mathsf T}
                      r\mathsf R_y(2\theta)]}{6}
 =\frac12+\frac r6[1+2\cos(2\theta-2\theta_0)].
 \label{eq:calibratedgatefidelity}
\end{equation}
This is the fidelity of the stated channel model, not a spectrum-only
certification for a general noisy qubit gate. At the correct angle it is
$(1+r)/2$, while Eq.~\eqref{eq:gaterate} supplies the entanglement produced
after the separately assumed ideal CNOT.

Spectral rotation certification \cite{Miklin2026} uses channel-eigenvalue
moduli and phases to bound qubit-gate quality with unknown SPAM under specific
channel assumptions. The explicit \(\pi/2\)-pulse protocol matches the Bell-endpoint
rotation up to a unitary axis change; a generic finite-\(\gamma\) angle requires
an adapted protocol. The spectrum in Eq.~\eqref{eq:PTM} alone does not identify
the rotation axis, initial-state alignment, a general qubit noise channel,
or the CNOT fidelity. Equation~\eqref{eq:gaterate} consequently assumes all of
these preparation details rather than claiming a black-box resource
certificate. Likewise, a nonunitary gauge transformation in gate-set
representations should not be confused with changing a physical state by
\(S_\gamma\) in a fixed inner product. Spectral or representation invariance
does not by itself imply identical physical preparation resources.

\section{Replica eigenmodes and the parity fixed-point algebra}
\label{app:replica}
\subsection{When a creation operator is exactly invariant}
Orthogonal fermionic replica mixing \cite{HoshinoFNG} acts on $q$ copies as
\begin{equation}
 \widehat W f_m^{(a)\dagger}\widehat W^\dagger
 =\sum_{b=1}^qW_{ba}f_m^{(b)\dagger},\qquad
 W^{\mathsf T}W=I_q,\quad\widehat W\ket{\mathrm{vac}}=\ket{\mathrm{vac}}.
 \label{eq:replicaW}
\end{equation}
The mode index $m$ is distinct from the replica index $a$. All creation
operators obey the canonical anticommutation relations across replicas.
For a normalized vector $w\in\mathbb C^q$, define
\begin{equation}
 g_m^\dagger(w)=\sum_aw_a f_m^{(a)\dagger},\qquad
 \{g_m(w),g_n^\dagger(w')\}=\delta_{mn}w^\dagger w'.
 \label{eq:collectivemode}
\end{equation}
Linearity of conjugation gives the exact calculation
\begin{align}
 \widehat W g_m^\dagger(w)\widehat W^\dagger
 &=\sum_{a,b}w_aW_{ba}f_m^{(b)\dagger}
 =\sum_b(Ww)_bf_m^{(b)\dagger}=g_m^\dagger(Ww).
 \label{eq:eigenmodeconjugation}
\end{align}
Creation operators are linearly independent: applying a proposed linear
relation to the vacuum gives a relation between orthonormal one-particle
states. Therefore $g_m^\dagger(w)$ is invariant if and only if $Ww=w$.
In the original replica basis this requires $We_a=e_a$; a fixed collective
mode does not imply that an individual original replica mode is fixed.

\subsection{Four-copy reflection and its three fixed difference modes}
For $v=(1,1,1,1)^{\mathsf T}/2$, the reflection is
\begin{equation}
 R=I_4-2vv^{\mathsf T}
 =\frac12\begin{pmatrix}
 1&-1&-1&-1\\-1&1&-1&-1\\-1&-1&1&-1\\-1&-1&-1&1
 \end{pmatrix},\quad R^2=I_4,\quad Rv=-v.
 \label{eq:reflectionmatrix}
\end{equation}
Every vector perpendicular to $v$ obeys $Rw=w$. An explicit orthonormal
basis of that three-dimensional space is
\begin{equation}
 w_1=\frac{(1,-1,0,0)^{\mathsf T}}{\sqrt2},\quad
 w_2=\frac{(1,1,-2,0)^{\mathsf T}}{\sqrt6},\quad
 w_3=\frac{(1,1,1,-3)^{\mathsf T}}{\sqrt{12}}.
 \label{eq:reflectiondifferencebasis}
\end{equation}
Put $g_{s,m}^\dagger=g_m^\dagger(v)$ and
$g_{j,m}^\dagger=g_m^\dagger(w_j)$. Equation~\eqref{eq:eigenmodeconjugation}
then becomes
\begin{equation}
 \widehat Rg_{s,m}^\dagger\widehat R^\dagger=-g_{s,m}^\dagger,
 \qquad\widehat Rg_{j,m}^\dagger\widehat R^\dagger=g_{j,m}^\dagger
 \quad(j=1,2,3).
 \label{eq:reflectionfixedmodes}
\end{equation}
Each original mode still mixes:
$\widehat R f_m^{(a)\dagger}\widehat R^\dagger
=f_m^{(a)\dagger}-g_{s,m}^\dagger$. None of the four columns is a
fixed original-basis column. With $n_{s,m}=g_{s,m}^\dagger g_{s,m}$,
an implementation fixing the vacuum is
\begin{equation}
 \widehat R=\prod_m(1-2n_{s,m}).
 \label{eq:reflectionFock}
\end{equation}
The identity $(1-2n_s)g_s^\dagger(1-2n_s)=-g_s^\dagger$ follows from
$n_sg_s^\dagger=g_s^\dagger$ and $g_s^\dagger n_s=0$; all orthogonal
mode operators commute with $n_s$. Thus Eq.~\eqref{eq:reflectionFock}
reproduces the conjugation rule, is Hermitian and unitary, and fixes the
vacuum. Symmetric-mode occupations are invariant despite the sign change
of their creation operators; Fock states acquire $(-1)^{\sum_m n_{s,m}}$.

\subsection{The common fixed-point principle and its precise scope}
For any Hermitian unitary $U$ with $P_\pm=(I\pm U)/2$, conjugation acts
as $+1$ on $P_+XP_+$ and $P_-XP_-$ and as $-1$ on the two cross blocks.
The random-unitary channel is therefore
\begin{equation}
 (1-q)X+qUXU
 =P_+XP_++P_-XP_-+(1-2q)(P_+XP_-+P_-XP_+).
 \label{eq:generalinvolutionchannel}
\end{equation}
For $q\ne0$, its fixed operators are exactly those commuting with $U$.
Density operators supported on either eigenspace are fixed; a vector may
acquire a minus global phase and still have an invariant density matrix.
Taking $U=Z_AZ_B$ gives Eq.~\eqref{eq:parityblocks}. Taking $U=\widehat R$
gives the analogous Fock-space fixed algebra, with
Eq.~\eqref{eq:reflectionfixedmodes} giving explicit invariant operators.
Both calculations use eigenspaces of a symmetry representation. They do
not identify an odd creation operator with a physical density matrix, nor
identify a replica rotation with the original two-qubit noise channel.
The $+1$ one-particle subspace is three-dimensional, whereas the adjoint
fixed algebra also includes products of two sign-flipped operators.

\subsection{Identical Gaussian inputs are invariant even when modes mix}
The original pure state supplies a further fixed-state calculation. For
$z=\sqrt{p_-/p_+}$, its $q$ identical fermionic copies have the normalized
BCS form
\begin{equation}
 \ket{\psi_\gamma}^{\otimes q}
 =p_+^{q/2}\exp\!\left(z\sum_{a=1}^q
 f_1^{(a)\dagger}f_2^{(a)\dagger}\right)\ket{\mathrm{vac}}.
 \label{eq:replicaBCS}
\end{equation}
The pair-creation factors are even and commute between copies, making this
exponential equivalent to the product of the $q$ one-copy expressions.
For real orthogonal $W$,
\begin{align}
 \widehat W\left(\sum_af_1^{(a)\dagger}f_2^{(a)\dagger}\right)\widehat W^\dagger
 &=\sum_{a,b,c}W_{ba}W_{ca}f_1^{(b)\dagger}f_2^{(c)\dagger}\notag\\
 &=\sum_{b,c}(WW^{\mathsf T})_{bc}f_1^{(b)\dagger}f_2^{(c)\dagger}
 =\sum_bf_1^{(b)\dagger}f_2^{(b)\dagger}.
 \label{eq:BCSreplicaproof}
\end{align}
The vacuum is fixed, hence the full state is fixed. This argument uses
orthogonality, not the existence of a $+1$ one-particle eigenmode. For example
a nontrivial two-copy rotation can have no eigenvalue $+1$ and still leave
Eq.~\eqref{eq:replicaBCS} invariant. In the convolution of
Eq.~\eqref{eq:FNGdefinition}, the invariant pure product has a pure one-copy
reduction, proving $\nu_2=1$. Generic non-Gaussian states do not have this
BCS invariant and still require the full replica trace calculation.

\end{document}